\documentclass[aps,pra,reprint,10pt]{revtex4-2}
\pdfoutput=1
\usepackage{braket}
\usepackage{mathtools}
\usepackage{adjustbox}
\usepackage[utf8]{inputenc}
\usepackage{graphicx}

\usepackage{csquotes}
\usepackage[dvipsnames]{xcolor}
\usepackage{color}
\usepackage{tikz}
\usepackage{graphics}
\usetikzlibrary{shapes.geometric,backgrounds,positioning,shapes.geometric,decorations.markings,decorations.pathreplacing,arrows,knots,hobby,angles,quotes}
\usepackage{amsmath}
\usepackage{amssymb}
\usepackage{amsthm}
\usepackage{braket}
\usepackage{bbm}
\usepackage{bm} 
\usepackage{siunitx}

\newcommand{\rubidium}{$^{87}$Rb }

\usepackage{hyperref}

\usepackage{cleveref}
\crefname{section}{section}{sections}
\Crefname{section}{Section}{Sections}
\crefname{equation}{Eq.}{Eq.}
\Crefname{equation}{Eq.}{Eq.}
\crefname{figure}{Fig.}{Fig.}
\Crefname{figure}{Fig.}{Fig.}
\crefname{appendix}{appendix}{appendices}
\Crefname{appendix}{Appendix}{Appendices}

\usepackage{pgfplots}
\pgfplotsset{compat=1.14}

\usepackage{amsfonts} 
\usepackage{lmodern,adjustbox}
\usepackage[skins,breakable]{tcolorbox}
\usepackage{tikz}
\usetikzlibrary{quantikz2}
\tcbuselibrary{listings}
\tcbuselibrary{breakable}
\usepackage{textcomp}

\usepackage{changepage}

\usetikzlibrary{decorations.markings}

\hypersetup{pdfauthor={Gabriel Müller, Victor J. Mart\'nez-Lahuerta, Ivan Sekulic, Sven Burger, Philipp-Immanuel Schneider, Naceur Gaaloul},pdftitle={Bayesian optimization for state engineering of quantum gases}}

\begin{document}

\title{Bayesian optimization for state engineering of quantum gases}

\author{Gabriel Müller$^1$}
\email{g.mueller@iqo.uni-hannover.de}
\author{V. J. Mart\'inez-Lahuerta$^1$}
\author{Ivan Sekulic$^{2,3}$}
\author{Sven Burger$^{2,3}$}
\author{Philipp-Immanuel Schneider$^{2,3}$}
\author{Naceur Gaaloul$^1$}

\affiliation{$^1$Institut f\"ur Quantenoptik, Leibniz Universit\"at Hannover, Germany}
\affiliation{$^2$JCMwave GmbH, Germany}
\affiliation{$^3$Zuse Institute Berlin, Germany}

\begin{abstract}
State engineering of quantum objects is a central requirement in most implementations. In the cases where the quantum dynamics can be described by analytical solutions or simple approximation models, optimal state preparation protocols have been theoretically proposed and experimentally realized.
For more complex systems, however, such as multi-component quantum gases, simplifying assumptions do not apply anymore and the optimization techniques become computationally impractical.
Here, we propose Bayesian optimization based on multi-output Gaussian processes to learn the quantum state's physical properties from few simulations only.
We evaluate its performance on an optimization study case of diabatically transporting a Bose-Einstein condensate while keeping it in its ground state, and show that within only few hundreds of executions of the underlying physics simulation, we reach a competitive performance with other protocols.
While restricting this benchmarking to well known approximations for straightforward comparisons, we expect a similar performance when employing more involving models, which are computationally more challenging. This paves the way to efficient state engineering of complex quantum systems.
\end{abstract}

\maketitle
\onecolumngrid

\section{Introduction}

\textbf{The accurate state engineering of quantum gases~\cite{Ketterle2002,Cornell2002} is a crucial requirement for performing fundamental physics experiments and for implementing them in real-world sensors.}
Fundamental physics experiments include the investigation of few-body physics~\cite{Naidon2017}, the preparation of entangled states~\cite{Corgier2021} and the search for new physics~\cite{Safronova2018}. Various other fields require well-controlled atomic states such as quantum simulations~\cite{Georgescu2014}, quantum communication~\cite{Gisin2007} or quantum sensing~\cite{Degen2017}. All of these applications rely on an efficient, robust and precise control and manipulation of quantum states to reach their ultimate potential in accuracy.

\textbf{In this work, we present a new approach of optimal state engineering based on Bayesian optimization and evaluate its performance in the case of preparing quantum gases for atom interferometry experiments.}
Atom interferometers are an important class of quantum sensors~\cite{Bongs2019a} with several applications, including the most precise determination of the fine-structure constant ~\cite{Parker2018a,Morel2020a}, tests of general relativity~\cite{Asenbaum2020,Ahlers2022}, geodesy~\cite{Menoret2018,Wu2019,Trimeche2019a,Leveque2023}, detection of gravitational waves~\cite{Graham2013,Canuel2018,Loriani2019,Zhan2020,Badurina2020,Canuel2020,Abe2021} and inertial sensing~\cite{Cheiney2018,Geiger2020,Hensel2021}. 
Here, the atoms' initial states and their uncertainties directly couple into the sensor performance and can cause systematic effects~\cite{Hensel2021,Struckmann2024}.
Therefore, these experiments require accurate protocols for state engineering to help overcome some of the current limitations~\cite{Asenbaum2020} and to reach unprecedented sensitivities enabling the study of new physics.

\textbf{For single atomic species, including Bose-Einstein condensates (BECs), state engineering protocols were successfully demonstrated.}
Optimally, these protocols enable a transport of the BEC with precise control over its full dynamics, made possible in some cases by a decoupled description of its center of mass (COM) and size dynamics~\cite{Qi2021}. 
The experimental implementation of the shortcut-to-adiabaticity (STA) protocol for instance achieved residual oscillations in a $\qty{0.93}{\milli \metre}$ displaced trap of only $0.4\pm \qty{0.15}{\micro \metre}$~\cite{Gaaloul2022}.
While STA shows a remarkable level of control over the COM dynamics, it cannot be used for controlling the size dynamics.
Still, with post-application of additional methods, the expansion energies of \rubidium BECs can be reduced to below $\qty{100}{\pico\kelvin}$ in 3D~\cite{Deppner2021a, Gaaloul2022}.
In order to include the size dynamics in the optimization, optimal control theory (OCT) protocols have been theoretically studied~\cite{Peirce1988, Jager2014}. Gradient-based approaches that gradually improve the dynamics starting from an initial transport scheme were recently proposed~\cite{Amri2019}.

\textbf{The currently existing BEC state engineering protocols are, however, not sufficient to prepare quantum states of more complex systems.}
They are mainly limited by their need for approximations to either find analytical solutions to the problem (STA) or to reduce the optimization's computational cost (OCT).
Such approximations include the simplification of trapping potentials to the harmonic case and the description of quantum dynamics only within the Thomas-Fermi regime~\cite{Castin1996}.
In particular, the OCT method described in~\cite{Amri2019}, requires up to millions of executions of the underlying physics simulation and therefore relies on these simplifications to be computationally feasible.
More complex systems such as the state engineering of two interacting BECs of different atomic species, greatly limit the validity of these approximations.
Yet, the insufficient control over the BECs is one of the factors, currently limiting the sensitivity of atom interferometry experiments on ground~\cite{Asenbaum2020} and in space~\cite{Elliott2023a}.
To control these dynamics accurately, one needs to employ computationally much more demanding simulations for directly solving the Gross-Pitaevskii equation (GPE)~\cite{Gross1963,Pitaevskii1961a,Pichery2023}.

\textbf{To engineer degenerate quantum states of matter by solving the GPE, one needs optimization methods that require significantly fewer executions of the physics simulation than what OCT does.}
Here, we report on using Bayesian optimization (BO)~\cite{Shahriari2016} with Gaussian processes (GPs) as a surrogate model~\cite{Rasmussen2006}. 
BO has been shown to require significantly fewer iterations than several other heuristic global optimization methods~\cite{Schneider2019, Plock2022, Anton2024}.
This feature makes BO especially interesting for problems where executions of the underlying process are costly in time or resources. 
In this work, we consider an extension of BO, that uses an advanced way of training a set of multi-output Gaussian processes to directly learn some of the quantum state's properties instead of only learning the final optimization goal as one scalar output.
In contrast to OCT, BO is a \emph{global} optimization method that does not rely on a good initial scheme, but rather on a suitable parametrization basis that can represent the desired transports.
We propose B-splines~\cite{deBoor1978} as an efficient but still flexible choice of a parametrization basis.

\textbf{Our implementation of Bayesian optimization finds competitive state engineering protocols in a few 100 executions of the underlying BEC dynamics simulation.}
We perform this optimization on a real-world example of a BEC transport relevant for many experiments~\cite{Gaaloul2022,Deppner2021a,Canuel2020,Trimeche2019a,Ahlers2022}.
To enable a comparison with the OCT results from~\cite{Amri2019}, we choose the same approximations, and, without this being an exact benchmark, we highlight that our approach requires only a fraction of the executions to realize competitive transports.
The significantly reduced number of required BEC simulations opens the path for using accurate modelling of complex systems during the optimization rather than relying on simple approximations such as the Thomas-Fermi one.
We further show that the extended BO method converges far better than classical BO.

\section{Methods}\label{section:methods}

\textbf{As benchmark for a state engineering problem, we choose the optimization goal of finding a quick BEC transport between two displaced traps with different strengths.}
We approximate the initial trapping potential harmonically via $\bm{\omega}^\text{i} \equiv \bm{\omega}\left(t=0\right) = 2\pi \times [f_x, f_y, f_z]$ with the trap origin at $x_0=y_0=0,\; z_0^\text{i}\equiv z_0\left(t=0\right)$.
Within some displacement time $t_f$, we move the trap smoothly in the z-direction such that the final trap is described by $\bm{\omega}^\text{f} \equiv \bm{\omega}(t=t_f)$ and $z_0^\text{f} \equiv z_0(t=t_f)$.
The BEC is initially trapped in its ground state and, ideally, ends up at the new ground state of the transported trap.
The optimization goal is to find a transport that minimizes the residual center-of-mass and size oscillations in the final trap.

\textbf{We implement trap properties typical for BEC experiments and already analysed with other optimization algorithms \cite{Corgier2018,Amri2019}.}
Such experiments use atom-chip generated magnetic traps \cite{Gaaloul2022,Deppner2021a,Becker2018} or optical dipole traps \cite{Weber2003,Herbst2024a} to create and manipulate BECs.
In particular, we use a magnetic trap configuration from \cite{Amri2019} to enable a comparison of our optimization method with previous OCT results.
The initial trap is located at $z^\text{i}_0\approx \qty{0.45}{\milli \metre}$ with trap frequencies $\bm{\omega}^\text{i} \approx 2 \pi \times [ 15, 615, 617] \; \unit{\hertz}$.
From there, within a given transport duration $\qty{50}{\milli\second} \le t_f \le \qty{200}{\milli\second}$, we transition into the final trap located at $z^\text{f}_0\approx \qty{1.65}{\milli \metre}$ with trap frequencies $\bm{\omega}^\text{f} \approx 2 \pi \times [ 10, 33, 31] \; \unit{\hertz}$.
Although such small transport durations result in non-adiabatic dynamics \cite{Masuda2009}, the realization of shortcuts to adiabaticity transports \cite{Torrontegui2013} in favor of adiabatic transports is usually desirable due to experimental constraints.
To control the transports shown here, we use a magnetic offset field $B\left(t\right)$ as a single time-dependent control parameter which can experimentally be realized by modifying a current flowing through a set of coils.
Due to the single control parameter, the trap frequencies are directly coupled to the trap's minimum position.

\textbf{We simulate the BEC transport dynamics within the Thomas-Fermi approximation, allowing for a direct comparison with the results of \cite{Amri2019}.}
The assumptions made here consist in neglecting both the BEC's initial kinetic energy and any anharmonic contributions to the trapping potential \cite{Castin1996,Pethick2002}.
This simplifies the description of the BEC dynamics to solving two independent sets of differential equations: \cref{eq:comEquations} describes the center of mass $z_A$ relying solely on Newton's equations and \cref{eq:sizeEquations} describes the 3-dimensional size dynamics using the so-called scaling approach of reference \cite{Castin1996}:
\begin{align}
    \ddot{z}_A\left(t\right) &= -\omega_z^2\left(t\right) \left( z_A\left(t\right) - z_0\left( t \right)\right), \label{eq:comEquations}\\
    \ddot{\lambda}_i\left(t\right) &= \frac{\omega_i^2\left(0\right)}{\lambda_i \left(t\right) \lambda_x \left(t\right) \lambda_y \left(t\right) \lambda_z \left(t\right)} - \omega_i^2\left(t\right) \lambda _i \left(t\right).\label{eq:sizeEquations}
\end{align}
Here, $\lambda_i$ for $i=x,y,z$ are scaling factors defining the time-dependent Thomas-Fermi radii $r_i\left( t \right) = \lambda_i \left(t\right) r_i\left(t=0\right)$ starting from the initial ground state, Thomas-Fermi radii $r_i\left(t=0\right)$ with initialization of the scaling factors as $\lambda_i=1$ and their time derivative $\dot {\lambda}_i=0$.
Both differential equations could be numerically integrated at a low computational cost.

\textbf{We express the optimization goal in terms of an objective function $C_\text{obj}$ that is minimized for desirable transports.}
The objective function includes the BEC's \textit{classical} energy $E_\text{cl}\left( t_f\right)$ and its \textit{quantum} energy $E_\text{qu}\left( t_f\right)$ at the end of the transport.
Minimizing these serves the purpose of reducing residual COM and size oscillations after the transport.
The respective definitions are:
\begin{align}
    E_\text{cl}\left( t \right) &= \frac{m}{2} \left( \omega_z^2[z_A-z_0]^2 + [\dot z_A-\dot z_0 ]^2 \right), \label{eq:classical_energy} \\
    E_\text{qu}(t)&=\frac{m}{14}\left[\omega_x^2 r_x^2+\omega_y^2 r_y^2+\omega_z^2 r_z^2\right]+\frac{m}{14}\left[\dot{r}_x^2+\dot{r}_y^2+\dot{r}_z^2\right]+\frac{15 g N}{28 \pi r_x r_y r_z}. \label{eq:quantum_energy}
\end{align}
with \rubidium mass $m$, scattering amplitude $g=4\pi \hbar ^2 a_s / m$, s-wave scattering length $a_s$ of \rubidium and atom number $N=\num{e5}$.
The \textit{classical} energy can be solely calculated from the center-of-mass dynamics whereas the \textit{quantum} energy is determined by the size dynamics.
An additional objective function contribution $E_\text{cl}^\text{int}$ quantifies the BEC's COM oscillations during the transport.
The motivation for minimizing $E_\text{cl}^\text{int}$ is to reduce the exploration of potential trap anharmonicities in an experimental realization of the transport:
\begin{align}
    E_\text{cl}^\text{int} = \frac{1}{t_f} \int _0 ^{t_f} E_\text{cl}\left(t\right). \label{eq:classical_energy_integral}
\end{align}
The full objective function $C_\text{obj}$ is aligned with the definition in \cite{Amri2019} and can be written as
\begin{align}
    C_\text{obj} = \lambda _\text{cl} E_\text{cl}\left(t_f\right) + \lambda _\text{qu} E_\text{qu}\left(t_f\right) + \lambda _\text{cl}^\text{int} E_\text{cl}^\text{int} ,\label{eq:optimisation_objective}
\end{align}
with positive real numbers for each of the weight factors.
Depending on the specific optimization goal, these weight factors can be used to balance the different magnitudes of the individual terms or drive the optimization focus.
Note that the \textit{quantum} energy $E_\text{qu}\left(t_f\right)$ is constrained from the bottom by its ground state energy in the final trap, defined here as $E_\text{qu}^0$.
It can be easily found by calculating the respective Thomas-Fermi radii for the final trap frequencies $\bm{\omega}^\text{f}$ and inserting those into \cref{eq:quantum_energy}.
Thus, with the other energy contributions being positive, the full objective function is subject to the constraint $C_\text{obj} \ge C_\text{obj}^0 \equiv \lambda_\text{qu} E_\text{qu}^0$.

\begin{figure}[htp]
    \centering
    \includegraphics[width=0.98\textwidth]{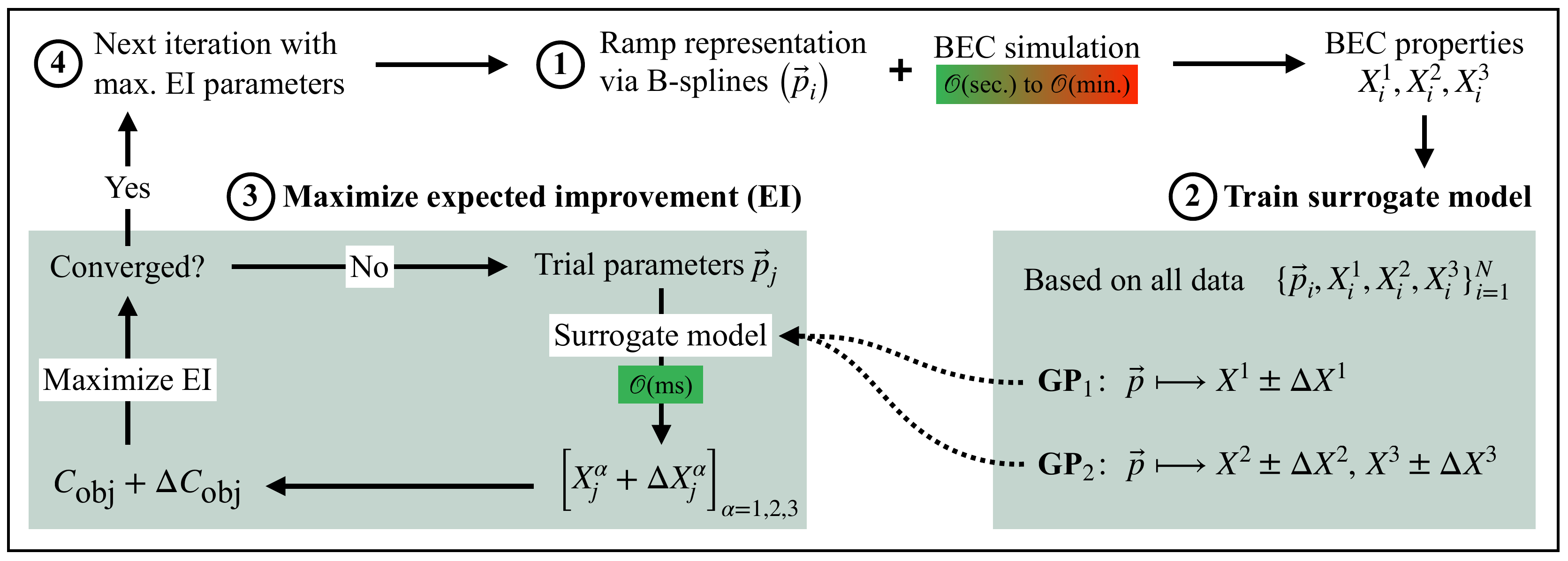}
    \caption{\textbf{Bayesian optimization (BO) with multi-output Gaussian processes (GPs) as a surrogate model.} The optimization process consists of four iterative steps: 1) generate a ramp representation via B-splines using parameters $\vec{p}_i$ and perform a BEC simulation with BEC properties as output, 2) based on all so far collected pair-wise data (parameters and BEC properties), a set of multi-output GPs is trained, 3) in a decoupled optimization loop, the surrogate model is used to explore the parameter landscape and maximize the expected improvement (EI), 4) the parameter set with maximized EI is used to restart the four-step loop. The advantage of using BO is that global exploration of the parameter landscape can be performed with a computationally cheap surrogate model instead of a potentially very expensive BEC simulation.}
    \label{fig:bo_gp_overview}
\end{figure}

\textbf{As an optimization method, we implement Bayesian optimization (BO) with Gaussian processes (GPs) as a surrogate model.}
The main idea of using this method is to replace some executions of the underlying BEC transport simulation (\textit{real} model) with a surrogate model \cite{Shahriari2016}.
Ideally, this surrogate model is much cheaper to evaluate than the \textit{real} model but still yields accurate predictions of the BEC dynamics.
As a surrogate model, we use GPs that we train with each evaluation of the \textit{real} model.
A GP is a stochastic process that learns to maps input parameters $\vec{p}_j$ to the appropriate output $X_j$ together with an input-dependent uncertainty $\Delta X_j$.
We can use the GP prediction on trial input parameters to calculate an expected improvement (EI) of the objective function $C_\text{obj}$ in comparison to previously observed \textit{real} model evaluations.
The EI can be large either due to small GP predictions of $C_\text{obj}$ with small uncertainty (exploitation) or due to large uncertainties in $C_\text{obj}$ that allow for improvement (exploration).
Using the duality of a \textit{real} model and a surrogate model, we perform an optimization loop (shown in \cref{fig:bo_gp_overview}) as follows: 1) execute the \textit{real} model with parameters $\vec{p}_i$ and add the resulting BEC properties $X_i^\alpha$ to our \textit{real} data collection, 2) train the surrogate model on all \textit{real} model input output pairs, 3) maximise the EI in the full parameter landscape by repeated evaluation of the surrogate model, 4) use the parameter set with maximum EI for the next \textit{real} model evaluation: $\vec{p}_{i+1} = \vec{p}_\text{max}^\text{ EI}$.
With this strategy, Bayesian optimization is a \textit{global} optimization method that balances the exploration of unknown transport candidates with an exploitation of promising transports.
It moves a majority of the computational effort of exploring the full parameter landscape with a potentially very costly \textit{real} model into the evaluation of a cheap surrogate model with fixed training cost.

\textbf{In order to guide the optimization process in an efficient way, we apply a monotonic transformation to the objective function.}
As described above, the EI strategy balances the convergence into local minima (exploitation) with the sampling at positions of large uncertainty (exploration). The relative importance given to exploitation and exploration depends on the scaling of the minimization objective. In our case, we target an optimal value of $C_\text{obj}$ that is only fractions of a \unit{\nano\kelvin} larger than its physical lower bound $C_\text{obj}^0=\lambda_\text{qu} E_\text{qu}^0$. For bad transports, $C_\text{obj}$ can take larger values by many orders of magnitude. Therefore, the convergence into local minima (exploitation) would already stop for $C_\text{obj} - C_\text{obj}^0 \approx \qty{1}{\nano\kelvin}$ in favor of an exploration of the parameter space. To give more emphasis on exploitation, we consider a monotonic transformation $g(C_\text{obj})$ that maps values close to $C_\text{obj}^0$ to a larger scale. A typical candidate transformation would be $g(C_\text{obj}) = \log\left(C_\text{obj} - C_\text{obj}^0\right)$. Within our approach, this transformation is problematic, however. Since the GPs do not encode the knowledge of a strictly positive \textit{quantum} energy, they can predict values $C_\text{obj} < C_\text{obj}^0$ leading to undefined values of the $\log$ function. Therefore, we use the alternative transformation $g(C_\text{obj}) =\operatorname{arsinh}\left(C_\text{obj} - C_\text{obj}^0\right)$ which has the strongest gradient for $C_\text{obj} = C_\text{obj}^0$ and is defined for all values of $C_\text{obj}$.

\textbf{In particular, we implement Bayesian optimization using a set of multi-output GPs.}
The GPs learn the mapping from transport ramp parameters to the values of the BEC properties $z_A, \dot z_A, r_x, \dot r_x, r_y, \dot r_y, r_z, \dot r_z$ at time $t=t_f$, as well as to the value of the integrated \textit{classical} energy $E_\text{cl}^\text{int}$. It is possible to learn each of the 9 variables with an independent scalar GP or to learn them with a single GP with 9 outputs. The latter approach is computationally more efficient, since only one covariance matrix needs to be inverted~\cite{Plock2022}. However, this approach disregards that each variable can have a different sensitivity towards the input parameters. While 9 independent GPs provide more accurate predictions, the computational cost increases approximately by a factor of 9. Therefore, we also consider an intermediate approach by grouping the variables into 3 multi-output GPs, where GP$_1$ learns the COM variables, GP$_2$ learns the size variables, and GP$_3$ learns $E_\text{cl}^\text{int}$.
In order to make predictions for the objective $C_\text{obj}$ and its uncertainty, we draw $N_\text{s} = \num{e4}$ samples from the GP's predicted normal distribution of the BEC properties. The corresponding $N_\text{s}$ values of $C_\text{obj}$ are used to determine the expected improvement by a Monte-Carlo integration \cite{Schneider2024}.

\textbf{Learning the underlying BEC properties has important advantages over classical Bayesian optimization.}
Classical BO trains a single scalar GP to map transport candidates directly to the scalar objective function value $C_\text{obj}$. However, this approach suffers from an information loss since the BEC dynamics is compressed into a single value. The final objective $C_\text{obj}$ depends on the squares and inverse values  of the BEC properties. Therefore, it can be harder to make predictions only based on previous observations of $C_\text{obj}$. For example, the value of $C_\text{obj}$ does not reveal direct information whether the final COM position $z_A\left(t_f\right)$ is smaller or larger than the targeted trap position $z_0\left(t_f\right)$. A direct prediction for values of $z_A\left(t_f\right)$ enables thus a more targeted correction of the transport aiming at decreasing the value of the \textit{classical} energy of Eq.~\eqref{eq:classical_energy}. Hence, learning the BEC properties can lead to a better convergence in the overall Bayesian optimization.

\textbf{For Bayesian optimization as a parametrized approach, we require an efficient basis to represent desired time-dependent transport ramps.}
There are requirements for the parametrization basis, given by both our optimization approach and the transport problem itself.
In general, the convergence of Bayesian optimization is deteriorated by too high-dimensional parameter spaces that make a global search inefficient.
Although, a priori, no specific maximal parameter space size can be defined, an efficient transport parametrization should use as few parameters as possible.
On the other side, the parametrization needs to be able to represent rather general time-dependent transport ramps to be able to reach the optimization goal as well as possible.
As explained above, the transport is defined by a single time-dependent experimental control parameter $B\left(t\right)$ that directly influences the trap properties.
In addition to the problem-defined initial and final value for the magnetic field $B^\text{i,f}$, to ensure experimental feasibility we also need vanishing first- and second-order time derivatives at the transport boundaries $\dot{B}^\text{i,f} = \ddot{B}^\text{i,f} = 0$.
To comply with experimental limitations, it is also helpful to require the parametrized ramp to be a function with smoothness of at least 2, i.e., it has a continuous second-order derivative over the full transport domain.

\textbf{We implement B-splines as a parametrization basis that fulfills all requirements to enable the desired transports.}
B-splines are functions that are composed of multiple piece-wise polynomial functions as described in detail in \cite{deBoor1978}.
A B-spline is uniquely defined by its order $n$, a set of knot points $\bm{k}$ and a set of control points $\bm{c}$.
Its order $n$ defines the order of the piece-wise polynomials to $n-1$ and consequently, allows for an overall smoothness of $n-2$.
Thus, in order to fulfill the smoothness requirement of at least 2, we assume $n\ge 4$.
The knot points $\bm{k}$ define the points in time at which the individual piece-wise polynomials are defined as well as neighbouring ones are connected.
By choosing the number of knot points, the total number of individual polynomials can be set.
The control points $\bm{c}$ define the shape of the B-spline as these are the values approached by the full B-spline function.
In order to fulfill the boundary conditions for the initial and final magnetic field values and its first- and second-order derivative, we apply two measures.
First, we set the first $\left(n-1\right)$ control points to $B^\text{i}$ and the final $\left(n-1\right)$ ones to $B^\text{f}$.
Additionally, we set the first $\left(n+1\right)$ knot points to $t=0$ and the final $\left(n+1\right)$ ones to $t=t_f$.
Now, we can use the remaining $N_c$ control points and $N_k$ knot points as our optimization parameters $\vec{p}$.
We make the choice of not optimising their absolute values directly but their relative values such that $c_{\left(n-1\right)+m} = \sum _{m=1}^{N_c} p_m$ and $k_{\left(n+1\right)+m} = \sum _{m=N_c+1}^{N_c+N_k} p_m$ with a final normalisation step omitted here.

\begin{figure*}[ht!]
    \centering
    \includegraphics[width=0.95\textwidth]{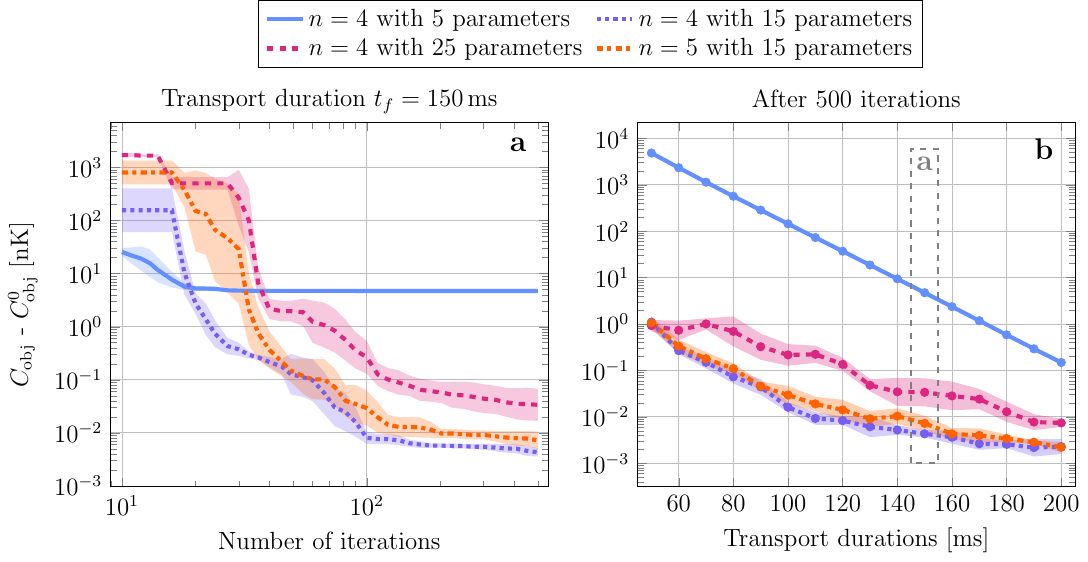}
    \caption{
    \textbf{Benchmark of different ramp parametrizations}.
    For different B-spline order and number of control and knot points, we perform 8 independent optimization runs and plot the average difference between the realized objective function value and its lower bound $C_\text{obj} - C_\text{obj}^0$ along with the standard deviation.
    We show four different kinds of B-splines, whereas on purpose, we include one under-parametrized one with only 5 free parameters and one over-parametrized one with 25 parameters.
    Panel \textbf{(a)} shows the convergence behavior for a transport duration of \qty{150}{\milli\second} for up to \num{500} iterations.
    BO converges best for a B-spline of order 4 with 15 free parameters. Panel \textbf{(b)} shows the performance after 500 iterations for different transport durations between \qty{50}{\milli\second} and \qty{200}{\milli\second}.
    The same B-spline parametrization with 15 parameters performs best overall.
    Parameters for this experiment: objective function weights $\lambda_\text{cl}=\num{1}, \lambda_\text{qu}=\num{3.3}, \lambda_\text{cl}^\text{int}=\num{5.5e-4}$, three multi-output GPs and scattering length $a_s=\num{98}a_0$ with Bohr radius $a_0$.
    }
    \label{fig:parametrisation_comparison}
\end{figure*}

\textbf{We test different kinds of B-spline parametrizations and find good convergence behavior with 15 free parameters.}
\Cref{fig:parametrisation_comparison} shows a collection of B-splines of order 4 and 5 with a different number of free parameters, respectively.
We evaluate the performance mainly in terms of two properties.
First, we check the convergence behavior for a \qty{150}{\milli\second} transport in terms of iterations.
Second, we probe their ability to minimize the objective function for different cases of transport durations. 
As an example of a too poorly resourced parametrization basis, we show a B-spline of order 4 with only 5 free parameters.
Over the full range of transport durations, it seems unable to represent
transports that reach competitive levels.
In contrast, by providing a B-spline of same order 4 a larger amount of 15 free parameters, it reaches much smaller final objective values within a feasible amount of iterations and for all transport duration cases.
We also show the optimization performance using a B-spline with 25 free parameters.
It clearly emphasises the worse convergence speed when using a parameter space larger than required.
Additionally, we show the influence of increasing the order of the B-spline to 5 while choosing the same well performing number of 15 parameters from before.
For this restricted number of 500 iterations analysed here, the overall performance does not improve.
In conclusion, for a specific transport problem, one could carry out a more detailed analysis and decide which parametrization basis to use based on a trade off between ramp flexibility and convergence requirements.
For the rest of this work, we use a B-spline parametrization of order 4 with 15 free parameters.

\section{Results}\label{section:results}

\textbf{We evaluate whether Bayesian optimization is a suitable approach for the state engineering of quantum gases based on the requirements of current precision experiments.}
In the context of this work, the requirements of leading BEC state engineering experiments can be expressed in terms of energy scales.
We highlight the energy scales on the order of tens of \unit{\pico \kelvin} experimentally realised in the uncertainty of COM dynamics \cite{Gaaloul2022} or in the size dynamics control \cite{Gaaloul2022,Deppner2021a}.
Therefore, we expect any helpful state engineering protocol to be able to operate on the level of \qty{e-3}{\nano \kelvin} for both the COM dynamics (expressed in $E_\text{cl}$) and size dynamics (expressed in $E_\text{qu}$).
In addition to the final performance level, a quick convergence of our approach is crucial for enabling the envisioned leap to using more accurate, and thus, more demanding BEC dynamics simulations.
Given that BO is a non-deterministic approach, ensuring reliable and reproducible performance is essential.
Therefore, in all results presented here, we show an average of 8 independent optimization runs along with their standard deviation.
To ensure some level of generality of the results presented here, we perform the optimization over a range of experimentally relevant transport durations between \qtylist{50;200}{\milli \second}.
In general, longer transport durations lead to more adiabatic transports, i.e., the state can remain closer to the immediate ground state of the system.
Thus, we expect better performance in terms of minimizing the objective function, Eq.~\eqref{eq:optimisation_objective}, for longer transport durations.
To perform detailed convergence analyses, we consistently use a transport duration of $\qty{150}{\milli \second}$ as a very typical choice in experiments.

\begin{figure*}[htp!]
    \centering
    \includegraphics[width=0.95\textwidth]{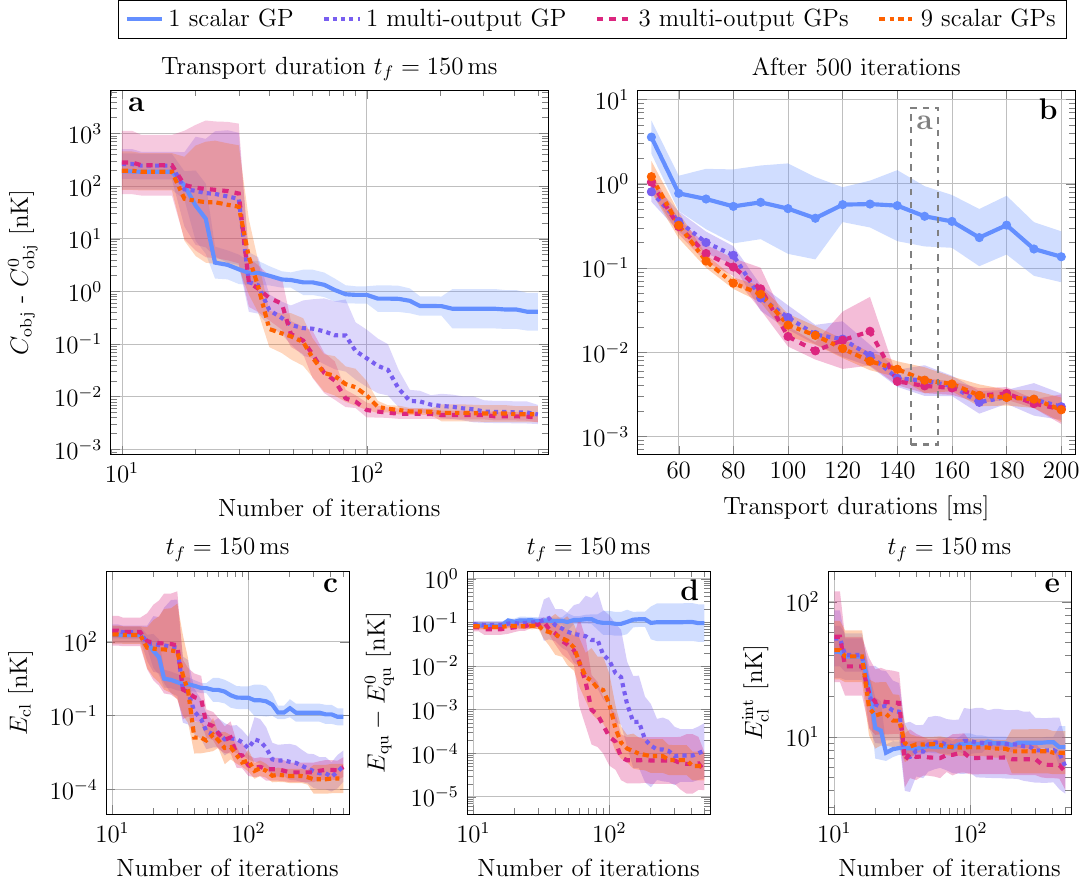}
    \caption{
    \textbf{Benchmark of classical BO vs. extended BO using differently separated multi-output GPs}.
    Instead of learning the objective value with one scalar GP (classical BO), our implementation learns 9 BEC properties that we can freely distribute over a variable number of multi-output GPs.
    Here, we show the performance for three different choices: learning all properties with one multi-output GP, distributing them over three multi-output GPs or learn all with 9 individual scalar GPs.
    For the data shown here, we performed 8 independent optimization runs and show their averages along with the standard deviation.
    Panel \textbf{(a)}, as well as panels \textbf{(c)}-\textbf{(e)} show the convergence behavior for a transport duration of \qty{150}{\milli\second} for up to \num{500} iterations.
    All extended BO implementations converge better than the classical one whereas the exact separation into different GPs has only a small impact.
    Panel \textbf{(b)} shows the performance after 500 iterations for different transport durations between \qtyrange{50}{200}{\milli\second}.
    Parameters for this experiment: objective function weights $\lambda_\text{cl}=\num{1}, \lambda_\text{qu}=\num{3.3}, \lambda_\text{cl}^\text{int}=\num{5.5e-4}$, B-spline of order 4 with 15 free parameters, and scattering length $a_s=\num{98}a_0$ with Bohr radius $a_0$.
    }
    \label{fig:multivariable_gp_comparison}
\end{figure*}

\textbf{The Bayesian optimization performance is greatly enhanced by using multi-output GPs and learning the BEC transport properties instead of just the scalar objective.}
In the classical implementation of BO, only one scalar GP is trained on the objective function value directly.
In complex and highly oscillatory parameter landscapes, it can be difficult for the optimizer to find the best transport from just this one scalar.
The performance of such a classical approach with one scalar GP is shown in \cref{fig:multivariable_gp_comparison}.
In comparison, we benchmark three different variants of learning the BEC dynamics: one multi-output GP learning all 9 BEC transport properties, three multi-output GPs each learning the relevant properties to the respective energy contributions in \cref{eq:optimisation_objective} and 9 scalar GPs each learning one of the properties.
All three variants find much better transports than their counterpart of just using a single scalar GP.
We observe this enhanced performance over the full spectrum of transport durations as shown in \cref{fig:multivariable_gp_comparison}b.
We assign this to a better understanding of the underlying BEC dynamics which enables the optimization on a simpler parameter landscape at the properties level instead of the objective function level.
The multi-output GP approach is particularly helpful in the convergence of the \textit{classical} and \textit{quantum} energies.
We assume this is due to the non-linear contributions of the BEC properties to these energies whereas the \textit{classical} integrated energy enters the objective function linearly.
The main advantage of using multi-output GPs instead of 9 individual scalar GPs for each property is computational efficiency.
Our benchmarks show no significant convergence difference between using three multi-output GPs compared to 9 individual scalar GPs.
However, in the intermediate convergence regime of around \num{100} iterations, we see a slight advantage of splitting the properties into three GPs instead of using just one GP with all outputs.
Therefore, in our case, the computationally most efficient choice would be to learn the BEC transport properties with three multi-output GPs.

\begin{figure}[htp]
    \centering
    \includegraphics[width=0.95\textwidth]{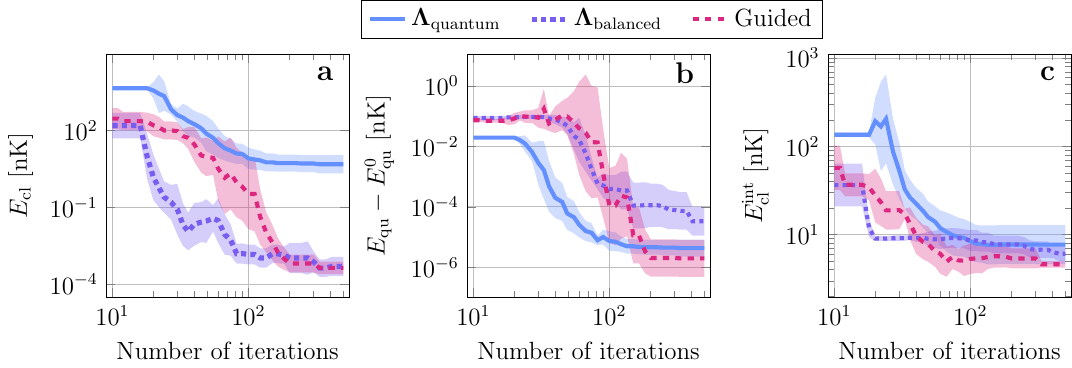}
    \caption{
        \textbf{Convergence behavior of objective function contributions for different optimization goals.}
        We evaluate the performance of three different optimizations goals, two based solely on weights taken over from benchmarks performed with OCT in this reference~\cite{Amri2019}, and one optimization guided by an absolute optimization goal.
        For the solely weight based optimizations, we use $\bm{\Lambda}_\text{balanced} = \left[\num{1}, \num{3.3}, \num{5.5d-4}\right]$ and $\bm{\Lambda}_\text{quantum} = \left[\num{1}, \num{5d5}, \num{1d-3}\right]$.
        For the guided case, we define the optimization goal $E_\text{cl} < \qty{e-3}{\nano \kelvin}$, $E_\text{qu} < E_\text{qu}^0 + \qty{e-5}{\nano \kelvin}$ and $E_\text{cl}^\text{int} < \qty{5}{\nano \kelvin}$.
        The panels \textbf{(a-c)} show the three objective function contributions, respectively.
        The guided optimization reaches the best performance in all contributions after $\sim \num{200}$ iterations. Before reaching \num{200} iterations, the balanced case achieves the smallest \textit{classical} energies while the quantum case achieves the smallest \textit{quantum} energies.
        Parameters for this experiment: transport duration of \qty{150}{\milli\second}, B-spline of order 4 with 15 free parameters, three multi-output GPs and scattering length $a_s=\num{99.6}a_0$ with Bohr radius $a_0$ (taken over from~\cite{Amri2019}).
    }
    \label{fig:convergence_optGoal}
\end{figure}

\textbf{We test the flexibility of our optimization approach by applying different weights to the individual objective function contributions.}
By setting the weights $\bm{\Lambda} = \left[\lambda_\text{cl}, \lambda_\text{qu}, \lambda_\text{cl}^\text{int} \right]$ in the objective function \cref{eq:optimisation_objective}, one indirectly defines an optimization goal.
The optimizer will primarily reduce the energy contributions with larger weights as long as their values change on the same order of magnitude.
We test two sets of weights taken over from \cite{Amri2019}, namely, $\bm{\Lambda}_\text{balanced} = \left[\num{1}, \num{3.3}, \num{5.5d-4}\right]$ and $\bm{\Lambda}_\text{quantum} = \left[\num{1}, \num{5d5}, \num{1d-3}\right]$.
Their convergence results for a $\qty{150}{\milli \second}$ transport are shown in \cref{fig:convergence_optGoal}.
We can clearly see how the convergence behavior is different for the two cases.
In the quantum-focused optimization, the \textit{quantum} energy approaches the ground state energy $E_\text{qu}^0$ much quicker compared to the balanced case.
This, however, impinges on the \textit{classical} energy which is not optimized as well as in the balanced case.
In both cases, the optimization seems to be converged for the most part after only $\sim \num{200}$ iterations.
After 500 iterations, the energies for $\bm{\Lambda}_\text{quantum}$ reach $E_\text{cl} \approx \qty{10}{\nano \kelvin}$, $E_\text{qu} \approx E_\text{qu}^0 + \qty{e-5}{\nano \kelvin}$, $E_\text{cl}^\text{int} \approx \qty{8}{\nano \kelvin}$ while for $\bm{\Lambda}_\text{balanced}$ they reach $E_\text{cl} \approx \qty{e-3}{\nano \kelvin}$, $E_\text{qu} \approx E_\text{qu}^0 + \qty{e-4}{\nano \kelvin}$, $E_\text{cl}^\text{int} \approx \qty{5}{\nano \kelvin}$.
The transport parametrization with B-splines seems to be able to represent ramps with residual \textit{classical} or \textit{quantum} energies below \unit{\pico \kelvin}.
However, the question remains if with a suitable choice of weights, the optimization can also find a single transport that reduces both energies to such levels simultaneously.

\textbf{To target an absolute optimization goal, we can guide the BO with additional constraints instead of solely relying on an appropriate choice of weights.}
In principle, a good starting point for reaching a desired ratio of different objective function contributions is to choose a matching ratio for the respective weights.
However, as this is only a relative definition of our optimization goal, the optimizer might get stuck in exploring options that compensate bad performance in one of the goals by overly optimizing another one.
In practice, this can end up in a tedious and expensive process of trial and error of different weights to counteract this behavior.
Therefore, it would be helpful to define absolute optimization goals in a way that avoids over-optimizing some contributions at the expense of others.
With BO, we can implement constraints as a means of guiding the optimization to reach our respective goal.
We follow the approach of \cite{Gardner2014} and adapt the internal evaluation of good parameter samples to maximising the product $\text{EI}\cdot\text{PF}$ with the expected improvement $\text{EI}$ and the probability of fulfilling the constraints $\text{PF}$.
We test the implementation of optimization constraints by trying to reach the combined optimal values for each of the energy contributions from the two previously tested weights $\bm{\Lambda}_\text{quantum}$ and $\bm{\Lambda}_\text{balanced}$.
These are $E_\text{cl} < \qty{e-3}{\nano \kelvin}$, $E_\text{qu} < E_\text{qu}^0 + \qty{e-5}{\nano \kelvin}$ and $E_\text{cl}^\text{int} < \qty{5}{\nano \kelvin}$.
From there, we normalise the weights such that their relative contribution after reaching the optimization goal stays equal.
We show the performance of this guided BO in \cref{fig:convergence_optGoal}.
Indeed, the guided optimization converges in all three different contributions in a rather balanced way, approaching the respective target performances similarly.
In addition to its balancing effect, the guided optimization even reaches the best overall performance in all contributions.

\textbf{Without this being an exact benchmark, our implementation of guided Bayesian optimization finds a competitive transport in a fraction of the BEC simulation executions compared to existing methods.}
In particular, for a $\qty{150}{\milli\second}$ transport, and within \num{200} iterations, we reach \textit{classical} energies on the order of \unit{\pico \kelvin} and residual deviations from the \textit{quantum} ground state energy well below the \unit{\pico \kelvin} level.
Compared to optimization results obtained with OCT \cite{Amri2019}, our convergence behavior reduces the required number of BEC simulations for reaching this performance level at least ten-fold.

\begin{figure}[htp]
    \centering
    \includegraphics[width=0.95\textwidth]{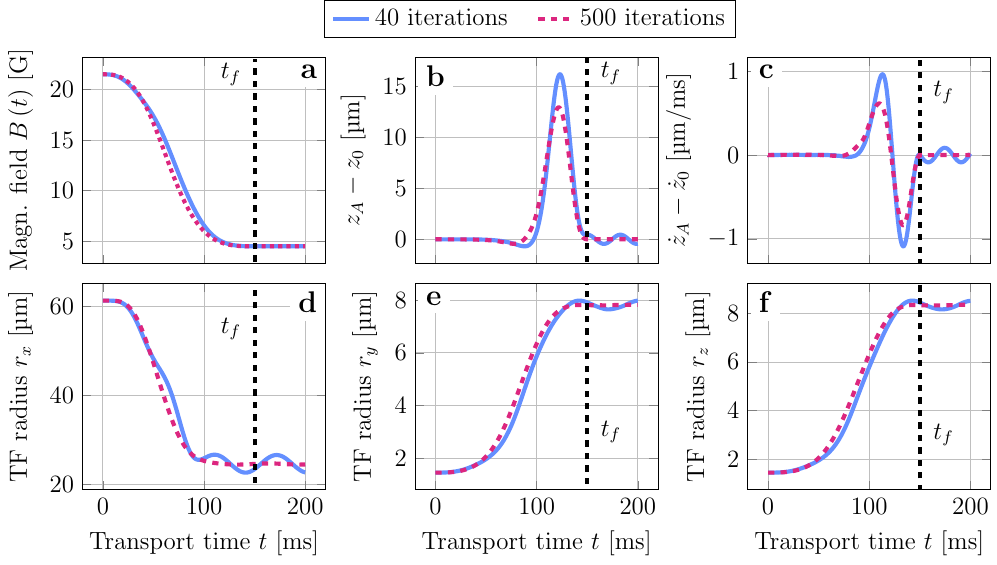}
    \caption{
    \textbf{Overview of BEC dynamics for two different transport tramps.}
    Two examples of a \qty{150}{\milli\second} transport, a not-optimal one after \num{40} iterations and a converged one after \num{500} iterations. Panel \textbf{(a)} shows the magnetic field value controlling the BEC trap.
    Panels \textbf{(b)} and \textbf{(c)} show the \textit{classical} COM dynamics in the co-moving trap frame.
    Panels \textbf{(d-f)} show the BEC's size dynamics in terms of the Thomas Fermi radii.
    Parameters for this experiment: B-spline of order 4 with 15 free parameters, three multi-output GPs and scattering length $a_s=\num{98}a_0$ with Bohr radius $a_0$.
    }
    \label{fig:transport_properties}
\end{figure}

\textbf{We confirm that the optimization indeed converges towards experimentally feasible transport ramps.}
In \cref{fig:transport_properties}, we show two examples of \qty{150}{\milli\second} transports, a not-optimal one after \num{40} iterations and a converged one after \num{500} iterations.
As imposed by using B-splines of order \num{4}, thus, continuous second order derivatives, the magnetic field ramp does not show any signs of experimental impracticability.
For a detailed analysis of experimental imperfections such as trap anharmonicities and their impact on the validity of using the Thomas-Fermi approximation, we refer the interested reader to \cite{Corgier2018}.
In our envisioned use of BO with more accurate models, such imperfections could be included in the optimization directly with their impact considered within the optimal ramp definition.
As expected due to the well-converged \textit{classical} and \textit{quantum} energies presented before, the optimized transport ramp reduces the residual BEC dynamics in the final trap to only small COM and size oscillations.

\section{Discussion and Outlook}\label{chap5}

\textbf{We introduced and validated Bayesian optimization with multi-output Gaussian processes as a new technique for optimizing BEC transport ramps.}
In order to benchmark its performance, we evaluated it on a well known and experimentally relevant playground, already implemented in state-of-the-art transport methods, namely shortcut-to-adiabaticity \cite{Corgier2018, Gaaloul2022} and optimal control theory \cite{Amri2019}.
Besides the exact trap properties, we also use the same means of approximating the BEC dynamics in a non-expensive way within the Thomas Fermi approximation.
Our benchmark results indicate major advantages of our Bayesian optimization approach.
Compared to the OCT performance \cite{Amri2019}, we require a fraction of executions of the underlying physics simulation until reaching competitive transports with residual \textit{classical} and \textit{quantum} energies on the level of $\unit{\pico\kelvin}$.
Our approach converges also significantly faster than classical Bayesian optimization using a single scalar Gaussian process. We attribute this advantage to the more accurate prediction of the final objective value stemming from the set of multi-output Gaussian processes that are trained directly on the underlying physical dynamics.

\textbf{We explain the mathematical implementation of our optimization approach in another article, currently in preparation \cite{Schneider2024}.}
There, among other things, we present a more detailed analysis of the computational expense of training the Gaussian processes and their utilization for exploring the parameter landscape.
Compared to other approaches, training the Gaussian processes leads to an overhead on the order of seconds that is paying off especially for computationally demanding problems.
Indeed, our envisioned application of this optimization method for more complex state engineering problems is exactly such a case, where the computational cost of the numerical simulations by far exceeds this overhead.

\textbf{The quick convergence behavior of our optimization approach opens up new possibilities for considering more accurate simulations and realistic experimental conditions.}
With Bayesian optimization as a black box approach, we can implement complex numerical simulations in a straightforward manner.
For the case of this work's benchmark problem of BEC transport optimizations, we can remove approximations usually done within the BEC simulations for this problem by directly solving the Gross-Pitaevskii equation in 3D.
Here, we can consider time-dependent potentials based on realistic experimental conditions without relying on any harmonic approximations or decoupling of the size dynamics from the center of mass dynamics. 
This additionally allows the extension of transports to the multi-dimensional or multiple-control parameter cases.
However, the execution of such numerical simulations can be highly demanding in computation time.
This prevents the use of such simulations in optimization approaches relying on thousands or even millions of executions.
However, the quick convergence of our approach on the order of a few 100 executions of the BEC dynamics simulation promises the feasibility of such extensions.

\textbf{We propose using our approach for the optimization of dual-species BEC transports to directly improve the sensitivity of quantum sensors.}
There is an immediate need for such transports to improve, for instance, the performance of dual-species atom interferometry experiments for fundamental physics tests \cite{Asenbaum2020,Elliott2023a}.
Here, one of the limiting factors is the insufficient control over both species' center of mass and size dynamics.
In contrast to single-species transports, as considered in our benchmarks, no reliable protocols for designing the transport of two interacting species are available yet.
However, using numerical methods for the simulation of dual-species transports \cite{Pichery2023}, one can accurately simulate and optimize these dynamics, too.
In order to directly improve the quantum sensor's performance, one could define an optimization goal that relates the relevant BEC wave-function properties after the transport with the sensor's sensitivity.
Here, we can greatly gain from the feature of our Gaussian processes to learn the underlying dynamics of the system instead of just the scalar objective function landscape.
It is possible to learn the physical dynamics once and exploit this knowledge for the optimization of various atom interferometry schemes with different explicit objective functions.

\textbf{Finally, we would like to highlight that our implementation of Bayesian optimization would benefit a multitude of communities with quantum state engineering challenges.}
Here, using accurate simulations of the complex quantum-mechanical systems as the basis for optimization protocols can enable major improvements in experimental performance. 
The only strong limitation of BO is the requirement of a parametrization that can efficiently represent the state engineering process.
Therefore, we can envision a useful application of BO with multi-output GPs in various fields where the performance of an experiment or sensor is defined by the accurate preparation and coherent manipulation of quantum states.
Such fields include quantum information processing \cite{Makhlin2001}, ultracold molecules generation \cite{Carr2009}, quantum simulation \cite{Georgescu2014} as well as quantum sensing \cite{Degen2017}.

\begin{acknowledgments}
We gratefully acknowledge helpful discussion with E. Charron, T. Estrampes, S. Seckmeyer and C. Struckmann. 
We additionally thank E. Charron for sharing data about previous work.
We thank W. Herr, H. Ahlers, S. Seckmeyer and E. Rasel for contributions to initiate this research direction.
This work was funded by the German Space Agency at the German Aerospace Center (Deutsche Raumfahrtagentur im Deutschen Zentrum für Luft- und Raumfahrt, DLR) with funds provided by the German Federal Ministry of Economic Affairs and Climate Action due to an enactment of the German Bundestag under Grants No. 50WM2245A (CAL-II), No. 50WM2253A/B (AI-Quadrat) and No. 50WM2263A (CARIOQA-GE) and the Berlin Mathematics Research Center MATH+ (EXC-2046/1, project ID: 390685689). NG acknowledges support from the Deutsche Forschungsgemeinschaft (German Research Foundation) under Germany’s Excellence Strategy (EXC-2123 QuantumFrontiers Grants No. 390837967), through CRC 1227 (DQ-mat) within Projects No. A05 and through the QuantERA 2021 co-funded project No. 499225223 (SQUEIS).
\end{acknowledgments}

\section*{Data availability}
The data used in this manuscript is available from the corresponding author upon reasonable request.
\section*{Competing interests}
All authors declare no competing interests.



\bibliographystyle{apsrev4-2}
\bibliography{BibFile}

\begin{thebibliography}{56}%
\makeatletter
\providecommand \@ifxundefined [1]{%
 \@ifx{#1\undefined}
}%
\providecommand \@ifnum [1]{%
 \ifnum #1\expandafter \@firstoftwo
 \else \expandafter \@secondoftwo
 \fi
}%
\providecommand \@ifx [1]{%
 \ifx #1\expandafter \@firstoftwo
 \else \expandafter \@secondoftwo
 \fi
}%
\providecommand \natexlab [1]{#1}%
\providecommand \enquote  [1]{``#1''}%
\providecommand \bibnamefont  [1]{#1}%
\providecommand \bibfnamefont [1]{#1}%
\providecommand \citenamefont [1]{#1}%
\providecommand \href@noop [0]{\@secondoftwo}%
\providecommand \href [0]{\begingroup \@sanitize@url \@href}%
\providecommand \@href[1]{\@@startlink{#1}\@@href}%
\providecommand \@@href[1]{\endgroup#1\@@endlink}%
\providecommand \@sanitize@url [0]{\catcode `\\12\catcode `\$12\catcode `\&12\catcode `\#12\catcode `\^12\catcode `\_12\catcode `\%12\relax}%
\providecommand \@@startlink[1]{}%
\providecommand \@@endlink[0]{}%
\providecommand \url  [0]{\begingroup\@sanitize@url \@url }%
\providecommand \@url [1]{\endgroup\@href {#1}{\urlprefix }}%
\providecommand \urlprefix  [0]{URL }%
\providecommand \Eprint [0]{\href }%
\providecommand \doibase [0]{https://doi.org/}%
\providecommand \selectlanguage [0]{\@gobble}%
\providecommand \bibinfo  [0]{\@secondoftwo}%
\providecommand \bibfield  [0]{\@secondoftwo}%
\providecommand \translation [1]{[#1]}%
\providecommand \BibitemOpen [0]{}%
\providecommand \bibitemStop [0]{}%
\providecommand \bibitemNoStop [0]{.\EOS\space}%
\providecommand \EOS [0]{\spacefactor3000\relax}%
\providecommand \BibitemShut  [1]{\csname bibitem#1\endcsname}%
\let\auto@bib@innerbib\@empty
\bibitem [{\citenamefont {Ketterle}(2002)}]{Ketterle2002}%
  \BibitemOpen
  \bibfield  {author} {\bibinfo {author} {\bibfnamefont {W.}~\bibnamefont {Ketterle}},\ }\href {https://doi.org/10.1103/RevModPhys.74.1131} {\bibfield  {journal} {\bibinfo  {journal} {Reviews of Modern Physics}\ }\textbf {\bibinfo {volume} {74}},\ \bibinfo {pages} {1131} (\bibinfo {year} {2002})}\BibitemShut {NoStop}%
\bibitem [{\citenamefont {Cornell}\ and\ \citenamefont {Wieman}(2002)}]{Cornell2002}%
  \BibitemOpen
  \bibfield  {author} {\bibinfo {author} {\bibfnamefont {E.~A.}\ \bibnamefont {Cornell}}\ and\ \bibinfo {author} {\bibfnamefont {C.~E.}\ \bibnamefont {Wieman}},\ }\href {https://doi.org/10.1103/RevModPhys.74.875} {\bibfield  {journal} {\bibinfo  {journal} {Reviews of Modern Physics}\ }\textbf {\bibinfo {volume} {74}},\ \bibinfo {pages} {875} (\bibinfo {year} {2002})}\BibitemShut {NoStop}%
\bibitem [{\citenamefont {Naidon}\ and\ \citenamefont {Endo}(2017)}]{Naidon2017}%
  \BibitemOpen
  \bibfield  {author} {\bibinfo {author} {\bibfnamefont {P.}~\bibnamefont {Naidon}}\ and\ \bibinfo {author} {\bibfnamefont {S.}~\bibnamefont {Endo}},\ }\href {https://doi.org/10.1088/1361-6633/aa50e8} {\bibfield  {journal} {\bibinfo  {journal} {Reports on Progress in Physics}\ }\textbf {\bibinfo {volume} {80}},\ \bibinfo {pages} {056001} (\bibinfo {year} {2017})}\BibitemShut {NoStop}%
\bibitem [{\citenamefont {Corgier}\ \emph {et~al.}(2021)\citenamefont {Corgier}, \citenamefont {Gaaloul}, \citenamefont {Smerzi},\ and\ \citenamefont {Pezz{\`e}}}]{Corgier2021}%
  \BibitemOpen
  \bibfield  {author} {\bibinfo {author} {\bibfnamefont {R.}~\bibnamefont {Corgier}}, \bibinfo {author} {\bibfnamefont {N.}~\bibnamefont {Gaaloul}}, \bibinfo {author} {\bibfnamefont {A.}~\bibnamefont {Smerzi}},\ and\ \bibinfo {author} {\bibfnamefont {L.}~\bibnamefont {Pezz{\`e}}},\ }\href {https://doi.org/10.1103/PhysRevLett.127.183401} {\bibfield  {journal} {\bibinfo  {journal} {Physical Review Letters}\ }\textbf {\bibinfo {volume} {127}},\ \bibinfo {pages} {183401} (\bibinfo {year} {2021})}\BibitemShut {NoStop}%
\bibitem [{\citenamefont {Safronova}\ \emph {et~al.}(2018)\citenamefont {Safronova}, \citenamefont {Budker}, \citenamefont {DeMille}, \citenamefont {Kimball}, \citenamefont {Derevianko},\ and\ \citenamefont {Clark}}]{Safronova2018}%
  \BibitemOpen
  \bibfield  {author} {\bibinfo {author} {\bibfnamefont {M.~S.}\ \bibnamefont {Safronova}}, \bibinfo {author} {\bibfnamefont {D.}~\bibnamefont {Budker}}, \bibinfo {author} {\bibfnamefont {D.}~\bibnamefont {DeMille}}, \bibinfo {author} {\bibfnamefont {D.~F.~J.}\ \bibnamefont {Kimball}}, \bibinfo {author} {\bibfnamefont {A.}~\bibnamefont {Derevianko}},\ and\ \bibinfo {author} {\bibfnamefont {C.~W.}\ \bibnamefont {Clark}},\ }\href {https://doi.org/10.1103/RevModPhys.90.025008} {\bibfield  {journal} {\bibinfo  {journal} {Reviews of Modern Physics}\ }\textbf {\bibinfo {volume} {90}},\ \bibinfo {pages} {025008} (\bibinfo {year} {2018})}\BibitemShut {NoStop}%
\bibitem [{\citenamefont {Georgescu}\ \emph {et~al.}(2014)\citenamefont {Georgescu}, \citenamefont {Ashhab},\ and\ \citenamefont {Nori}}]{Georgescu2014}%
  \BibitemOpen
  \bibfield  {author} {\bibinfo {author} {\bibfnamefont {I.~M.}\ \bibnamefont {Georgescu}}, \bibinfo {author} {\bibfnamefont {S.}~\bibnamefont {Ashhab}},\ and\ \bibinfo {author} {\bibfnamefont {F.}~\bibnamefont {Nori}},\ }\href {https://doi.org/10.1103/RevModPhys.86.153} {\bibfield  {journal} {\bibinfo  {journal} {Reviews of Modern Physics}\ }\textbf {\bibinfo {volume} {86}},\ \bibinfo {pages} {153} (\bibinfo {year} {2014})}\BibitemShut {NoStop}%
\bibitem [{\citenamefont {Gisin}\ and\ \citenamefont {Thew}(2007)}]{Gisin2007}%
  \BibitemOpen
  \bibfield  {author} {\bibinfo {author} {\bibfnamefont {N.}~\bibnamefont {Gisin}}\ and\ \bibinfo {author} {\bibfnamefont {R.}~\bibnamefont {Thew}},\ }\href {https://doi.org/10.1038/nphoton.2007.22} {\bibfield  {journal} {\bibinfo  {journal} {Nature Photonics}\ }\textbf {\bibinfo {volume} {1}},\ \bibinfo {pages} {165} (\bibinfo {year} {2007})}\BibitemShut {NoStop}%
\bibitem [{\citenamefont {Degen}\ \emph {et~al.}(2017)\citenamefont {Degen}, \citenamefont {Reinhard},\ and\ \citenamefont {Cappellaro}}]{Degen2017}%
  \BibitemOpen
  \bibfield  {author} {\bibinfo {author} {\bibfnamefont {C.~L.}\ \bibnamefont {Degen}}, \bibinfo {author} {\bibfnamefont {F.}~\bibnamefont {Reinhard}},\ and\ \bibinfo {author} {\bibfnamefont {P.}~\bibnamefont {Cappellaro}},\ }\href {https://doi.org/10.1103/RevModPhys.89.035002} {\bibfield  {journal} {\bibinfo  {journal} {Reviews of Modern Physics}\ }\textbf {\bibinfo {volume} {89}},\ \bibinfo {pages} {035002} (\bibinfo {year} {2017})}\BibitemShut {NoStop}%
\bibitem [{\citenamefont {Bongs}\ \emph {et~al.}(2019)\citenamefont {Bongs}, \citenamefont {Holynski}, \citenamefont {Vovrosh}, \citenamefont {Bouyer}, \citenamefont {Condon}, \citenamefont {Rasel}, \citenamefont {Schubert}, \citenamefont {Schleich},\ and\ \citenamefont {Roura}}]{Bongs2019a}%
  \BibitemOpen
  \bibfield  {author} {\bibinfo {author} {\bibfnamefont {K.}~\bibnamefont {Bongs}}, \bibinfo {author} {\bibfnamefont {M.}~\bibnamefont {Holynski}}, \bibinfo {author} {\bibfnamefont {J.}~\bibnamefont {Vovrosh}}, \bibinfo {author} {\bibfnamefont {P.}~\bibnamefont {Bouyer}}, \bibinfo {author} {\bibfnamefont {G.}~\bibnamefont {Condon}}, \bibinfo {author} {\bibfnamefont {E.}~\bibnamefont {Rasel}}, \bibinfo {author} {\bibfnamefont {C.}~\bibnamefont {Schubert}}, \bibinfo {author} {\bibfnamefont {W.~P.}\ \bibnamefont {Schleich}},\ and\ \bibinfo {author} {\bibfnamefont {A.}~\bibnamefont {Roura}},\ }\href {https://doi.org/10.1038/s42254-019-0117-4} {\bibfield  {journal} {\bibinfo  {journal} {Nature Reviews Physics}\ }\textbf {\bibinfo {volume} {1}},\ \bibinfo {pages} {731} (\bibinfo {year} {2019})}\BibitemShut {NoStop}%
\bibitem [{\citenamefont {Parker}\ \emph {et~al.}(2018)\citenamefont {Parker}, \citenamefont {Yu}, \citenamefont {Zhong}, \citenamefont {Estey},\ and\ \citenamefont {M{\"u}ller}}]{Parker2018a}%
  \BibitemOpen
  \bibfield  {author} {\bibinfo {author} {\bibfnamefont {R.~H.}\ \bibnamefont {Parker}}, \bibinfo {author} {\bibfnamefont {C.}~\bibnamefont {Yu}}, \bibinfo {author} {\bibfnamefont {W.}~\bibnamefont {Zhong}}, \bibinfo {author} {\bibfnamefont {B.}~\bibnamefont {Estey}},\ and\ \bibinfo {author} {\bibfnamefont {H.}~\bibnamefont {M{\"u}ller}},\ }\href {https://doi.org/10.1126/science.aap7706} {\bibfield  {journal} {\bibinfo  {journal} {Science}\ }\textbf {\bibinfo {volume} {360}},\ \bibinfo {pages} {191} (\bibinfo {year} {2018})}\BibitemShut {NoStop}%
\bibitem [{\citenamefont {Morel}\ \emph {et~al.}(2020)\citenamefont {Morel}, \citenamefont {Yao}, \citenamefont {Clad{\'e}},\ and\ \citenamefont {{Guellati-Kh{\'e}lifa}}}]{Morel2020a}%
  \BibitemOpen
  \bibfield  {author} {\bibinfo {author} {\bibfnamefont {L.}~\bibnamefont {Morel}}, \bibinfo {author} {\bibfnamefont {Z.}~\bibnamefont {Yao}}, \bibinfo {author} {\bibfnamefont {P.}~\bibnamefont {Clad{\'e}}},\ and\ \bibinfo {author} {\bibfnamefont {S.}~\bibnamefont {{Guellati-Kh{\'e}lifa}}},\ }\href {https://doi.org/10.1038/s41586-020-2964-7} {\bibfield  {journal} {\bibinfo  {journal} {Nature}\ }\textbf {\bibinfo {volume} {588}},\ \bibinfo {pages} {61} (\bibinfo {year} {2020})}\BibitemShut {NoStop}%
\bibitem [{\citenamefont {Asenbaum}\ \emph {et~al.}(2020)\citenamefont {Asenbaum}, \citenamefont {Overstreet}, \citenamefont {Kim}, \citenamefont {Curti},\ and\ \citenamefont {Kasevich}}]{Asenbaum2020}%
  \BibitemOpen
  \bibfield  {author} {\bibinfo {author} {\bibfnamefont {P.}~\bibnamefont {Asenbaum}}, \bibinfo {author} {\bibfnamefont {C.}~\bibnamefont {Overstreet}}, \bibinfo {author} {\bibfnamefont {M.}~\bibnamefont {Kim}}, \bibinfo {author} {\bibfnamefont {J.}~\bibnamefont {Curti}},\ and\ \bibinfo {author} {\bibfnamefont {M.~A.}\ \bibnamefont {Kasevich}},\ }\href {https://doi.org/10.1103/PhysRevLett.125.191101} {\bibfield  {journal} {\bibinfo  {journal} {Physical Review Letters}\ }\textbf {\bibinfo {volume} {125}},\ \bibinfo {pages} {191101} (\bibinfo {year} {2020})}\BibitemShut {NoStop}%
\bibitem [{\citenamefont {Ahlers}\ \emph {et~al.}(2022)\citenamefont {Ahlers}, \citenamefont {Badurina}, \citenamefont {Bassi}, \citenamefont {Battelier}, \citenamefont {Beaufils}, \citenamefont {Bongs}, \citenamefont {Bouyer}, \citenamefont {Braxmaier}, \citenamefont {Buchmueller}, \citenamefont {Carlesso}, \citenamefont {Charron}, \citenamefont {Chiofalo}, \citenamefont {Corgier}, \citenamefont {Donadi}, \citenamefont {Droz}, \citenamefont {Ecoffet}, \citenamefont {Ellis}, \citenamefont {Est{\`e}ve}, \citenamefont {Gaaloul}, \citenamefont {Gerardi}, \citenamefont {Giese}, \citenamefont {Grosse}, \citenamefont {Hees}, \citenamefont {Hensel}, \citenamefont {Herr}, \citenamefont {Jetzer}, \citenamefont {Kleinsteinberg}, \citenamefont {Klempt}, \citenamefont {Lecomte}, \citenamefont {Lopes}, \citenamefont {Loriani}, \citenamefont {M{\'e}tris}, \citenamefont {Martin}, \citenamefont {Mart{\'i}n}, \citenamefont {M{\"u}ller}, \citenamefont {Nofrarias}, \citenamefont {Santos}, \citenamefont {Rasel}, \citenamefont
  {Robert}, \citenamefont {Saks}, \citenamefont {Salter}, \citenamefont {Schlippert}, \citenamefont {Schubert}, \citenamefont {Schuldt}, \citenamefont {Sopuerta}, \citenamefont {Struckmann}, \citenamefont {Tino}, \citenamefont {Valenzuela}, \citenamefont {{von Klitzing}}, \citenamefont {W{\"o}rner}, \citenamefont {Wolf}, \citenamefont {Yu},\ and\ \citenamefont {Zelan}}]{Ahlers2022}%
  \BibitemOpen
  \bibfield  {author} {\bibinfo {author} {\bibfnamefont {H.}~\bibnamefont {Ahlers}}, \bibinfo {author} {\bibfnamefont {L.}~\bibnamefont {Badurina}}, \bibinfo {author} {\bibfnamefont {A.}~\bibnamefont {Bassi}}, \bibinfo {author} {\bibfnamefont {B.}~\bibnamefont {Battelier}}, \bibinfo {author} {\bibfnamefont {Q.}~\bibnamefont {Beaufils}}, \bibinfo {author} {\bibfnamefont {K.}~\bibnamefont {Bongs}}, \bibinfo {author} {\bibfnamefont {P.}~\bibnamefont {Bouyer}}, \bibinfo {author} {\bibfnamefont {C.}~\bibnamefont {Braxmaier}}, \bibinfo {author} {\bibfnamefont {O.}~\bibnamefont {Buchmueller}}, \bibinfo {author} {\bibfnamefont {M.}~\bibnamefont {Carlesso}}, \bibinfo {author} {\bibfnamefont {E.}~\bibnamefont {Charron}}, \bibinfo {author} {\bibfnamefont {M.~L.}\ \bibnamefont {Chiofalo}}, \bibinfo {author} {\bibfnamefont {R.}~\bibnamefont {Corgier}}, \bibinfo {author} {\bibfnamefont {S.}~\bibnamefont {Donadi}}, \bibinfo {author} {\bibfnamefont {F.}~\bibnamefont {Droz}}, \bibinfo {author} {\bibfnamefont {R.}~\bibnamefont
  {Ecoffet}}, \bibinfo {author} {\bibfnamefont {J.}~\bibnamefont {Ellis}}, \bibinfo {author} {\bibfnamefont {F.}~\bibnamefont {Est{\`e}ve}}, \bibinfo {author} {\bibfnamefont {N.}~\bibnamefont {Gaaloul}}, \bibinfo {author} {\bibfnamefont {D.}~\bibnamefont {Gerardi}}, \bibinfo {author} {\bibfnamefont {E.}~\bibnamefont {Giese}}, \bibinfo {author} {\bibfnamefont {J.}~\bibnamefont {Grosse}}, \bibinfo {author} {\bibfnamefont {A.}~\bibnamefont {Hees}}, \bibinfo {author} {\bibfnamefont {T.}~\bibnamefont {Hensel}}, \bibinfo {author} {\bibfnamefont {W.}~\bibnamefont {Herr}}, \bibinfo {author} {\bibfnamefont {P.}~\bibnamefont {Jetzer}}, \bibinfo {author} {\bibfnamefont {G.}~\bibnamefont {Kleinsteinberg}}, \bibinfo {author} {\bibfnamefont {C.}~\bibnamefont {Klempt}}, \bibinfo {author} {\bibfnamefont {S.}~\bibnamefont {Lecomte}}, \bibinfo {author} {\bibfnamefont {L.}~\bibnamefont {Lopes}}, \bibinfo {author} {\bibfnamefont {S.}~\bibnamefont {Loriani}}, \bibinfo {author} {\bibfnamefont {G.}~\bibnamefont {M{\'e}tris}},
  \bibinfo {author} {\bibfnamefont {T.}~\bibnamefont {Martin}}, \bibinfo {author} {\bibfnamefont {V.}~\bibnamefont {Mart{\'i}n}}, \bibinfo {author} {\bibfnamefont {G.}~\bibnamefont {M{\"u}ller}}, \bibinfo {author} {\bibfnamefont {M.}~\bibnamefont {Nofrarias}}, \bibinfo {author} {\bibfnamefont {F.~P.~D.}\ \bibnamefont {Santos}}, \bibinfo {author} {\bibfnamefont {E.~M.}\ \bibnamefont {Rasel}}, \bibinfo {author} {\bibfnamefont {A.}~\bibnamefont {Robert}}, \bibinfo {author} {\bibfnamefont {N.}~\bibnamefont {Saks}}, \bibinfo {author} {\bibfnamefont {M.}~\bibnamefont {Salter}}, \bibinfo {author} {\bibfnamefont {D.}~\bibnamefont {Schlippert}}, \bibinfo {author} {\bibfnamefont {C.}~\bibnamefont {Schubert}}, \bibinfo {author} {\bibfnamefont {T.}~\bibnamefont {Schuldt}}, \bibinfo {author} {\bibfnamefont {C.~F.}\ \bibnamefont {Sopuerta}}, \bibinfo {author} {\bibfnamefont {C.}~\bibnamefont {Struckmann}}, \bibinfo {author} {\bibfnamefont {G.~M.}\ \bibnamefont {Tino}}, \bibinfo {author} {\bibfnamefont {T.}~\bibnamefont
  {Valenzuela}}, \bibinfo {author} {\bibfnamefont {W.}~\bibnamefont {{von Klitzing}}}, \bibinfo {author} {\bibfnamefont {L.}~\bibnamefont {W{\"o}rner}}, \bibinfo {author} {\bibfnamefont {P.}~\bibnamefont {Wolf}}, \bibinfo {author} {\bibfnamefont {N.}~\bibnamefont {Yu}},\ and\ \bibinfo {author} {\bibfnamefont {M.}~\bibnamefont {Zelan}},\ }\href {https://doi.org/10.48550/arXiv.2211.15412} {\bibinfo {title} {{{STE-QUEST}}: {{Space Time Explorer}} and {{QUantum Equivalence}} principle {{Space Test}}}} (\bibinfo {year} {2022}),\ \Eprint {https://arxiv.org/abs/2211.15412} {arxiv:2211.15412 [gr-qc, physics:hep-ex, physics:hep-ph, physics:physics, physics:quant-ph]} \BibitemShut {NoStop}%
\bibitem [{\citenamefont {M{\'e}noret}\ \emph {et~al.}(2018)\citenamefont {M{\'e}noret}, \citenamefont {Vermeulen}, \citenamefont {Le~Moigne}, \citenamefont {Bonvalot}, \citenamefont {Bouyer}, \citenamefont {Landragin},\ and\ \citenamefont {Desruelle}}]{Menoret2018}%
  \BibitemOpen
  \bibfield  {author} {\bibinfo {author} {\bibfnamefont {V.}~\bibnamefont {M{\'e}noret}}, \bibinfo {author} {\bibfnamefont {P.}~\bibnamefont {Vermeulen}}, \bibinfo {author} {\bibfnamefont {N.}~\bibnamefont {Le~Moigne}}, \bibinfo {author} {\bibfnamefont {S.}~\bibnamefont {Bonvalot}}, \bibinfo {author} {\bibfnamefont {P.}~\bibnamefont {Bouyer}}, \bibinfo {author} {\bibfnamefont {A.}~\bibnamefont {Landragin}},\ and\ \bibinfo {author} {\bibfnamefont {B.}~\bibnamefont {Desruelle}},\ }\href {https://doi.org/10.1038/s41598-018-30608-1} {\bibfield  {journal} {\bibinfo  {journal} {Scientific Reports}\ }\textbf {\bibinfo {volume} {8}},\ \bibinfo {pages} {12300} (\bibinfo {year} {2018})}\BibitemShut {NoStop}%
\bibitem [{\citenamefont {Wu}\ \emph {et~al.}(2019)\citenamefont {Wu}, \citenamefont {Pagel}, \citenamefont {Malek}, \citenamefont {Nguyen}, \citenamefont {Zi}, \citenamefont {Scheirer},\ and\ \citenamefont {M{\"u}ller}}]{Wu2019}%
  \BibitemOpen
  \bibfield  {author} {\bibinfo {author} {\bibfnamefont {X.}~\bibnamefont {Wu}}, \bibinfo {author} {\bibfnamefont {Z.}~\bibnamefont {Pagel}}, \bibinfo {author} {\bibfnamefont {B.~S.}\ \bibnamefont {Malek}}, \bibinfo {author} {\bibfnamefont {T.~H.}\ \bibnamefont {Nguyen}}, \bibinfo {author} {\bibfnamefont {F.}~\bibnamefont {Zi}}, \bibinfo {author} {\bibfnamefont {D.~S.}\ \bibnamefont {Scheirer}},\ and\ \bibinfo {author} {\bibfnamefont {H.}~\bibnamefont {M{\"u}ller}},\ }\href {https://doi.org/10.1126/sciadv.aax0800} {\bibfield  {journal} {\bibinfo  {journal} {Science Advances}\ }\textbf {\bibinfo {volume} {5}},\ \bibinfo {pages} {eaax0800} (\bibinfo {year} {2019})}\BibitemShut {NoStop}%
\bibitem [{\citenamefont {Trimeche}\ \emph {et~al.}(2019)\citenamefont {Trimeche}, \citenamefont {Battelier}, \citenamefont {Becker}, \citenamefont {Bertoldi}, \citenamefont {Bouyer}, \citenamefont {Braxmaier}, \citenamefont {Charron}, \citenamefont {Corgier}, \citenamefont {Cornelius}, \citenamefont {Douch}, \citenamefont {Gaaloul}, \citenamefont {Herrmann}, \citenamefont {M{\"u}ller}, \citenamefont {Rasel}, \citenamefont {Schubert}, \citenamefont {Wu},\ and\ \citenamefont {dos Santos}}]{Trimeche2019a}%
  \BibitemOpen
  \bibfield  {author} {\bibinfo {author} {\bibfnamefont {A.}~\bibnamefont {Trimeche}}, \bibinfo {author} {\bibfnamefont {B.}~\bibnamefont {Battelier}}, \bibinfo {author} {\bibfnamefont {D.}~\bibnamefont {Becker}}, \bibinfo {author} {\bibfnamefont {A.}~\bibnamefont {Bertoldi}}, \bibinfo {author} {\bibfnamefont {P.}~\bibnamefont {Bouyer}}, \bibinfo {author} {\bibfnamefont {C.}~\bibnamefont {Braxmaier}}, \bibinfo {author} {\bibfnamefont {E.}~\bibnamefont {Charron}}, \bibinfo {author} {\bibfnamefont {R.}~\bibnamefont {Corgier}}, \bibinfo {author} {\bibfnamefont {M.}~\bibnamefont {Cornelius}}, \bibinfo {author} {\bibfnamefont {K.}~\bibnamefont {Douch}}, \bibinfo {author} {\bibfnamefont {N.}~\bibnamefont {Gaaloul}}, \bibinfo {author} {\bibfnamefont {S.}~\bibnamefont {Herrmann}}, \bibinfo {author} {\bibfnamefont {J.}~\bibnamefont {M{\"u}ller}}, \bibinfo {author} {\bibfnamefont {E.}~\bibnamefont {Rasel}}, \bibinfo {author} {\bibfnamefont {C.}~\bibnamefont {Schubert}}, \bibinfo {author} {\bibfnamefont
  {H.}~\bibnamefont {Wu}},\ and\ \bibinfo {author} {\bibfnamefont {F.~P.}\ \bibnamefont {dos Santos}},\ }\href {https://doi.org/10.1088/1361-6382/ab4548} {\bibfield  {journal} {\bibinfo  {journal} {Classical and Quantum Gravity}\ }\textbf {\bibinfo {volume} {36}},\ \bibinfo {pages} {215004} (\bibinfo {year} {2019})}\BibitemShut {NoStop}%
\bibitem [{\citenamefont {L{\'e}v{\`e}que}\ \emph {et~al.}(2023)\citenamefont {L{\'e}v{\`e}que}, \citenamefont {Fallet}, \citenamefont {Lefebve}, \citenamefont {Piquereau}, \citenamefont {Gauguet}, \citenamefont {Battelier}, \citenamefont {Bouyer}, \citenamefont {Gaaloul}, \citenamefont {Lachmann}, \citenamefont {Piest}, \citenamefont {Rasel}, \citenamefont {M{\"u}ller}, \citenamefont {Schubert}, \citenamefont {Beaufils},\ and\ \citenamefont {Santos}}]{Leveque2023}%
  \BibitemOpen
  \bibfield  {author} {\bibinfo {author} {\bibfnamefont {T.}~\bibnamefont {L{\'e}v{\`e}que}}, \bibinfo {author} {\bibfnamefont {C.}~\bibnamefont {Fallet}}, \bibinfo {author} {\bibfnamefont {J.}~\bibnamefont {Lefebve}}, \bibinfo {author} {\bibfnamefont {A.}~\bibnamefont {Piquereau}}, \bibinfo {author} {\bibfnamefont {A.}~\bibnamefont {Gauguet}}, \bibinfo {author} {\bibfnamefont {B.}~\bibnamefont {Battelier}}, \bibinfo {author} {\bibfnamefont {P.}~\bibnamefont {Bouyer}}, \bibinfo {author} {\bibfnamefont {N.}~\bibnamefont {Gaaloul}}, \bibinfo {author} {\bibfnamefont {M.}~\bibnamefont {Lachmann}}, \bibinfo {author} {\bibfnamefont {B.}~\bibnamefont {Piest}}, \bibinfo {author} {\bibfnamefont {E.}~\bibnamefont {Rasel}}, \bibinfo {author} {\bibfnamefont {J.}~\bibnamefont {M{\"u}ller}}, \bibinfo {author} {\bibfnamefont {C.}~\bibnamefont {Schubert}}, \bibinfo {author} {\bibfnamefont {Q.}~\bibnamefont {Beaufils}},\ and\ \bibinfo {author} {\bibfnamefont {F.~P.~D.}\ \bibnamefont {Santos}},\ }in\ \href
  {https://doi.org/10.1117/12.2690536} {\emph {\bibinfo {booktitle} {International {{Conference}} on {{Space Optics}} --- {{ICSO}} 2022}}},\ Vol.\ \bibinfo {volume} {12777}\ (\bibinfo  {publisher} {SPIE},\ \bibinfo {year} {2023})\ pp.\ \bibinfo {pages} {1536--1545}\BibitemShut {NoStop}%
\bibitem [{\citenamefont {Graham}\ \emph {et~al.}(2013)\citenamefont {Graham}, \citenamefont {Hogan}, \citenamefont {Kasevich},\ and\ \citenamefont {Rajendran}}]{Graham2013}%
  \BibitemOpen
  \bibfield  {author} {\bibinfo {author} {\bibfnamefont {P.~W.}\ \bibnamefont {Graham}}, \bibinfo {author} {\bibfnamefont {J.~M.}\ \bibnamefont {Hogan}}, \bibinfo {author} {\bibfnamefont {M.~A.}\ \bibnamefont {Kasevich}},\ and\ \bibinfo {author} {\bibfnamefont {S.}~\bibnamefont {Rajendran}},\ }\href {https://doi.org/10.1103/PhysRevLett.110.171102} {\bibfield  {journal} {\bibinfo  {journal} {Physical Review Letters}\ }\textbf {\bibinfo {volume} {110}},\ \bibinfo {pages} {171102} (\bibinfo {year} {2013})}\BibitemShut {NoStop}%
\bibitem [{\citenamefont {Canuel}\ \emph {et~al.}(2018)\citenamefont {Canuel}, \citenamefont {Bertoldi}, \citenamefont {Amand}, \citenamefont {{Pozzo di Borgo}}, \citenamefont {Chantrait}, \citenamefont {Danquigny}, \citenamefont {Dovale~{\'A}lvarez}, \citenamefont {Fang}, \citenamefont {Freise}, \citenamefont {Geiger}, \citenamefont {Gillot}, \citenamefont {Henry}, \citenamefont {Hinderer}, \citenamefont {Holleville}, \citenamefont {Junca}, \citenamefont {Lef{\`e}vre}, \citenamefont {Merzougui}, \citenamefont {Mielec}, \citenamefont {Monfret}, \citenamefont {Pelisson}, \citenamefont {Prevedelli}, \citenamefont {Reynaud}, \citenamefont {Riou}, \citenamefont {Rogister}, \citenamefont {Rosat}, \citenamefont {Cormier}, \citenamefont {Landragin}, \citenamefont {Chaibi}, \citenamefont {Gaffet},\ and\ \citenamefont {Bouyer}}]{Canuel2018}%
  \BibitemOpen
  \bibfield  {author} {\bibinfo {author} {\bibfnamefont {B.}~\bibnamefont {Canuel}}, \bibinfo {author} {\bibfnamefont {A.}~\bibnamefont {Bertoldi}}, \bibinfo {author} {\bibfnamefont {L.}~\bibnamefont {Amand}}, \bibinfo {author} {\bibfnamefont {E.}~\bibnamefont {{Pozzo di Borgo}}}, \bibinfo {author} {\bibfnamefont {T.}~\bibnamefont {Chantrait}}, \bibinfo {author} {\bibfnamefont {C.}~\bibnamefont {Danquigny}}, \bibinfo {author} {\bibfnamefont {M.}~\bibnamefont {Dovale~{\'A}lvarez}}, \bibinfo {author} {\bibfnamefont {B.}~\bibnamefont {Fang}}, \bibinfo {author} {\bibfnamefont {A.}~\bibnamefont {Freise}}, \bibinfo {author} {\bibfnamefont {R.}~\bibnamefont {Geiger}}, \bibinfo {author} {\bibfnamefont {J.}~\bibnamefont {Gillot}}, \bibinfo {author} {\bibfnamefont {S.}~\bibnamefont {Henry}}, \bibinfo {author} {\bibfnamefont {J.}~\bibnamefont {Hinderer}}, \bibinfo {author} {\bibfnamefont {D.}~\bibnamefont {Holleville}}, \bibinfo {author} {\bibfnamefont {J.}~\bibnamefont {Junca}}, \bibinfo {author} {\bibfnamefont
  {G.}~\bibnamefont {Lef{\`e}vre}}, \bibinfo {author} {\bibfnamefont {M.}~\bibnamefont {Merzougui}}, \bibinfo {author} {\bibfnamefont {N.}~\bibnamefont {Mielec}}, \bibinfo {author} {\bibfnamefont {T.}~\bibnamefont {Monfret}}, \bibinfo {author} {\bibfnamefont {S.}~\bibnamefont {Pelisson}}, \bibinfo {author} {\bibfnamefont {M.}~\bibnamefont {Prevedelli}}, \bibinfo {author} {\bibfnamefont {S.}~\bibnamefont {Reynaud}}, \bibinfo {author} {\bibfnamefont {I.}~\bibnamefont {Riou}}, \bibinfo {author} {\bibfnamefont {Y.}~\bibnamefont {Rogister}}, \bibinfo {author} {\bibfnamefont {S.}~\bibnamefont {Rosat}}, \bibinfo {author} {\bibfnamefont {E.}~\bibnamefont {Cormier}}, \bibinfo {author} {\bibfnamefont {A.}~\bibnamefont {Landragin}}, \bibinfo {author} {\bibfnamefont {W.}~\bibnamefont {Chaibi}}, \bibinfo {author} {\bibfnamefont {S.}~\bibnamefont {Gaffet}},\ and\ \bibinfo {author} {\bibfnamefont {P.}~\bibnamefont {Bouyer}},\ }\href {https://doi.org/10.1038/s41598-018-32165-z} {\bibfield  {journal} {\bibinfo  {journal}
  {Scientific Reports}\ }\textbf {\bibinfo {volume} {8}},\ \bibinfo {pages} {14064} (\bibinfo {year} {2018})}\BibitemShut {NoStop}%
\bibitem [{\citenamefont {Loriani}\ \emph {et~al.}(2019)\citenamefont {Loriani}, \citenamefont {Schlippert}, \citenamefont {Schubert}, \citenamefont {Abend}, \citenamefont {Ahlers}, \citenamefont {Ertmer}, \citenamefont {Rudolph}, \citenamefont {Hogan}, \citenamefont {Kasevich}, \citenamefont {Rasel},\ and\ \citenamefont {Gaaloul}}]{Loriani2019}%
  \BibitemOpen
  \bibfield  {author} {\bibinfo {author} {\bibfnamefont {S.}~\bibnamefont {Loriani}}, \bibinfo {author} {\bibfnamefont {D.}~\bibnamefont {Schlippert}}, \bibinfo {author} {\bibfnamefont {C.}~\bibnamefont {Schubert}}, \bibinfo {author} {\bibfnamefont {S.}~\bibnamefont {Abend}}, \bibinfo {author} {\bibfnamefont {H.}~\bibnamefont {Ahlers}}, \bibinfo {author} {\bibfnamefont {W.}~\bibnamefont {Ertmer}}, \bibinfo {author} {\bibfnamefont {J.}~\bibnamefont {Rudolph}}, \bibinfo {author} {\bibfnamefont {J.~M.}\ \bibnamefont {Hogan}}, \bibinfo {author} {\bibfnamefont {M.~A.}\ \bibnamefont {Kasevich}}, \bibinfo {author} {\bibfnamefont {E.~M.}\ \bibnamefont {Rasel}},\ and\ \bibinfo {author} {\bibfnamefont {N.}~\bibnamefont {Gaaloul}},\ }\href {https://doi.org/10.1088/1367-2630/ab22d0} {\bibfield  {journal} {\bibinfo  {journal} {New Journal of Physics}\ }\textbf {\bibinfo {volume} {21}},\ \bibinfo {pages} {063030} (\bibinfo {year} {2019})}\BibitemShut {NoStop}%
\bibitem [{\citenamefont {Zhan}\ \emph {et~al.}(2020)\citenamefont {Zhan}, \citenamefont {Wang}, \citenamefont {Ni}, \citenamefont {Gao}, \citenamefont {Wang}, \citenamefont {He}, \citenamefont {Li}, \citenamefont {Zhou}, \citenamefont {Chen}, \citenamefont {Zhong}, \citenamefont {Tang}, \citenamefont {Yao}, \citenamefont {Zhu}, \citenamefont {Xiong}, \citenamefont {Lu}, \citenamefont {Yu}, \citenamefont {Cheng}, \citenamefont {Liu}, \citenamefont {Liang}, \citenamefont {Xu}, \citenamefont {He}, \citenamefont {Ke}, \citenamefont {Tan},\ and\ \citenamefont {Luo}}]{Zhan2020}%
  \BibitemOpen
  \bibfield  {author} {\bibinfo {author} {\bibfnamefont {M.-S.}\ \bibnamefont {Zhan}}, \bibinfo {author} {\bibfnamefont {J.}~\bibnamefont {Wang}}, \bibinfo {author} {\bibfnamefont {W.-T.}\ \bibnamefont {Ni}}, \bibinfo {author} {\bibfnamefont {D.-F.}\ \bibnamefont {Gao}}, \bibinfo {author} {\bibfnamefont {G.}~\bibnamefont {Wang}}, \bibinfo {author} {\bibfnamefont {L.-X.}\ \bibnamefont {He}}, \bibinfo {author} {\bibfnamefont {R.-B.}\ \bibnamefont {Li}}, \bibinfo {author} {\bibfnamefont {L.}~\bibnamefont {Zhou}}, \bibinfo {author} {\bibfnamefont {X.}~\bibnamefont {Chen}}, \bibinfo {author} {\bibfnamefont {J.-Q.}\ \bibnamefont {Zhong}}, \bibinfo {author} {\bibfnamefont {B.}~\bibnamefont {Tang}}, \bibinfo {author} {\bibfnamefont {Z.-W.}\ \bibnamefont {Yao}}, \bibinfo {author} {\bibfnamefont {L.}~\bibnamefont {Zhu}}, \bibinfo {author} {\bibfnamefont {Z.-Y.}\ \bibnamefont {Xiong}}, \bibinfo {author} {\bibfnamefont {S.-B.}\ \bibnamefont {Lu}}, \bibinfo {author} {\bibfnamefont {G.-H.}\ \bibnamefont {Yu}}, \bibinfo
  {author} {\bibfnamefont {Q.-F.}\ \bibnamefont {Cheng}}, \bibinfo {author} {\bibfnamefont {M.}~\bibnamefont {Liu}}, \bibinfo {author} {\bibfnamefont {Y.-R.}\ \bibnamefont {Liang}}, \bibinfo {author} {\bibfnamefont {P.}~\bibnamefont {Xu}}, \bibinfo {author} {\bibfnamefont {X.-D.}\ \bibnamefont {He}}, \bibinfo {author} {\bibfnamefont {M.}~\bibnamefont {Ke}}, \bibinfo {author} {\bibfnamefont {Z.}~\bibnamefont {Tan}},\ and\ \bibinfo {author} {\bibfnamefont {J.}~\bibnamefont {Luo}},\ }\href {https://doi.org/10.1142/S0218271819400054} {\bibfield  {journal} {\bibinfo  {journal} {International Journal of Modern Physics D}\ }\textbf {\bibinfo {volume} {29}},\ \bibinfo {pages} {1940005} (\bibinfo {year} {2020})}\BibitemShut {NoStop}%
\bibitem [{\citenamefont {Badurina}\ \emph {et~al.}(2020)\citenamefont {Badurina}, \citenamefont {Bentine}, \citenamefont {Blas}, \citenamefont {Bongs}, \citenamefont {Bortoletto}, \citenamefont {Bowcock}, \citenamefont {Bridges}, \citenamefont {Bowden}, \citenamefont {Buchmueller}, \citenamefont {Burrage}, \citenamefont {Coleman}, \citenamefont {Elertas}, \citenamefont {Ellis}, \citenamefont {Foot}, \citenamefont {Gibson}, \citenamefont {Haehnelt}, \citenamefont {Harte}, \citenamefont {Hedges}, \citenamefont {Hobson}, \citenamefont {Holynski}, \citenamefont {Jones}, \citenamefont {Langlois}, \citenamefont {Lellouch}, \citenamefont {Lewicki}, \citenamefont {Maiolino}, \citenamefont {Majewski}, \citenamefont {Malik}, \citenamefont {{March-Russell}}, \citenamefont {McCabe}, \citenamefont {Newbold}, \citenamefont {Sauer}, \citenamefont {Schneider}, \citenamefont {Shipsey}, \citenamefont {Singh}, \citenamefont {Uchida}, \citenamefont {Valenzuela}, \citenamefont {van~der Grinten}, \citenamefont {Vaskonen},
  \citenamefont {Vossebeld}, \citenamefont {Weatherill},\ and\ \citenamefont {Wilmut}}]{Badurina2020}%
  \BibitemOpen
  \bibfield  {author} {\bibinfo {author} {\bibfnamefont {L.}~\bibnamefont {Badurina}}, \bibinfo {author} {\bibfnamefont {E.}~\bibnamefont {Bentine}}, \bibinfo {author} {\bibfnamefont {D.}~\bibnamefont {Blas}}, \bibinfo {author} {\bibfnamefont {K.}~\bibnamefont {Bongs}}, \bibinfo {author} {\bibfnamefont {D.}~\bibnamefont {Bortoletto}}, \bibinfo {author} {\bibfnamefont {T.}~\bibnamefont {Bowcock}}, \bibinfo {author} {\bibfnamefont {K.}~\bibnamefont {Bridges}}, \bibinfo {author} {\bibfnamefont {W.}~\bibnamefont {Bowden}}, \bibinfo {author} {\bibfnamefont {O.}~\bibnamefont {Buchmueller}}, \bibinfo {author} {\bibfnamefont {C.}~\bibnamefont {Burrage}}, \bibinfo {author} {\bibfnamefont {J.}~\bibnamefont {Coleman}}, \bibinfo {author} {\bibfnamefont {G.}~\bibnamefont {Elertas}}, \bibinfo {author} {\bibfnamefont {J.}~\bibnamefont {Ellis}}, \bibinfo {author} {\bibfnamefont {C.}~\bibnamefont {Foot}}, \bibinfo {author} {\bibfnamefont {V.}~\bibnamefont {Gibson}}, \bibinfo {author} {\bibfnamefont {M.~G.}\ \bibnamefont
  {Haehnelt}}, \bibinfo {author} {\bibfnamefont {T.}~\bibnamefont {Harte}}, \bibinfo {author} {\bibfnamefont {S.}~\bibnamefont {Hedges}}, \bibinfo {author} {\bibfnamefont {R.}~\bibnamefont {Hobson}}, \bibinfo {author} {\bibfnamefont {M.}~\bibnamefont {Holynski}}, \bibinfo {author} {\bibfnamefont {T.}~\bibnamefont {Jones}}, \bibinfo {author} {\bibfnamefont {M.}~\bibnamefont {Langlois}}, \bibinfo {author} {\bibfnamefont {S.}~\bibnamefont {Lellouch}}, \bibinfo {author} {\bibfnamefont {M.}~\bibnamefont {Lewicki}}, \bibinfo {author} {\bibfnamefont {R.}~\bibnamefont {Maiolino}}, \bibinfo {author} {\bibfnamefont {P.}~\bibnamefont {Majewski}}, \bibinfo {author} {\bibfnamefont {S.}~\bibnamefont {Malik}}, \bibinfo {author} {\bibfnamefont {J.}~\bibnamefont {{March-Russell}}}, \bibinfo {author} {\bibfnamefont {C.}~\bibnamefont {McCabe}}, \bibinfo {author} {\bibfnamefont {D.}~\bibnamefont {Newbold}}, \bibinfo {author} {\bibfnamefont {B.}~\bibnamefont {Sauer}}, \bibinfo {author} {\bibfnamefont {U.}~\bibnamefont
  {Schneider}}, \bibinfo {author} {\bibfnamefont {I.}~\bibnamefont {Shipsey}}, \bibinfo {author} {\bibfnamefont {Y.}~\bibnamefont {Singh}}, \bibinfo {author} {\bibfnamefont {M.~A.}\ \bibnamefont {Uchida}}, \bibinfo {author} {\bibfnamefont {T.}~\bibnamefont {Valenzuela}}, \bibinfo {author} {\bibfnamefont {M.}~\bibnamefont {van~der Grinten}}, \bibinfo {author} {\bibfnamefont {V.}~\bibnamefont {Vaskonen}}, \bibinfo {author} {\bibfnamefont {J.}~\bibnamefont {Vossebeld}}, \bibinfo {author} {\bibfnamefont {D.}~\bibnamefont {Weatherill}},\ and\ \bibinfo {author} {\bibfnamefont {I.}~\bibnamefont {Wilmut}},\ }\href {https://doi.org/10.1088/1475-7516/2020/05/011} {\bibfield  {journal} {\bibinfo  {journal} {Journal of Cosmology and Astroparticle Physics}\ }\textbf {\bibinfo {volume} {2020}}\bibinfo  {number} { (05)},\ \bibinfo {pages} {011}}\BibitemShut {NoStop}%
\bibitem [{\citenamefont {Canuel}\ \emph {et~al.}(2020)\citenamefont {Canuel}, \citenamefont {Abend}, \citenamefont {{Amaro-Seoane}}, \citenamefont {Badaracco}, \citenamefont {Beaufils}, \citenamefont {Bertoldi}, \citenamefont {Bongs}, \citenamefont {Bouyer}, \citenamefont {Braxmaier}, \citenamefont {Chaibi}, \citenamefont {Christensen}, \citenamefont {Fitzek}, \citenamefont {Flouris}, \citenamefont {Gaaloul}, \citenamefont {Gaffet}, \citenamefont {Alzar}, \citenamefont {Geiger}, \citenamefont {{Guellati-Khelifa}}, \citenamefont {Hammerer}, \citenamefont {Harms}, \citenamefont {Hinderer}, \citenamefont {Holynski}, \citenamefont {Junca}, \citenamefont {Katsanevas}, \citenamefont {Klempt}, \citenamefont {Kozanitis}, \citenamefont {Krutzik}, \citenamefont {Landragin}, \citenamefont {Roche}, \citenamefont {Leykauf}, \citenamefont {Lien}, \citenamefont {Loriani}, \citenamefont {Merlet}, \citenamefont {Merzougui}, \citenamefont {Nofrarias}, \citenamefont {Papadakos}, \citenamefont {dos Santos}, \citenamefont
  {Peters}, \citenamefont {Plexousakis}, \citenamefont {Prevedelli}, \citenamefont {Rasel}, \citenamefont {Rogister}, \citenamefont {Rosat}, \citenamefont {Roura}, \citenamefont {Sabulsky}, \citenamefont {Schkolnik}, \citenamefont {Schlippert}, \citenamefont {Schubert}, \citenamefont {Sidorenkov}, \citenamefont {Siem{\ss}}, \citenamefont {Sopuerta}, \citenamefont {Sorrentino}, \citenamefont {Struckmann}, \citenamefont {Tino}, \citenamefont {Tsagkatakis}, \citenamefont {Vicer{\'e}}, \citenamefont {von Klitzing}, \citenamefont {Woerner},\ and\ \citenamefont {Zou}}]{Canuel2020}%
  \BibitemOpen
\bibfield  {number} {  }\bibfield  {author} {\bibinfo {author} {\bibfnamefont {B.}~\bibnamefont {Canuel}}, \bibinfo {author} {\bibfnamefont {S.}~\bibnamefont {Abend}}, \bibinfo {author} {\bibfnamefont {P.}~\bibnamefont {{Amaro-Seoane}}}, \bibinfo {author} {\bibfnamefont {F.}~\bibnamefont {Badaracco}}, \bibinfo {author} {\bibfnamefont {Q.}~\bibnamefont {Beaufils}}, \bibinfo {author} {\bibfnamefont {A.}~\bibnamefont {Bertoldi}}, \bibinfo {author} {\bibfnamefont {K.}~\bibnamefont {Bongs}}, \bibinfo {author} {\bibfnamefont {P.}~\bibnamefont {Bouyer}}, \bibinfo {author} {\bibfnamefont {C.}~\bibnamefont {Braxmaier}}, \bibinfo {author} {\bibfnamefont {W.}~\bibnamefont {Chaibi}}, \bibinfo {author} {\bibfnamefont {N.}~\bibnamefont {Christensen}}, \bibinfo {author} {\bibfnamefont {F.}~\bibnamefont {Fitzek}}, \bibinfo {author} {\bibfnamefont {G.}~\bibnamefont {Flouris}}, \bibinfo {author} {\bibfnamefont {N.}~\bibnamefont {Gaaloul}}, \bibinfo {author} {\bibfnamefont {S.}~\bibnamefont {Gaffet}}, \bibinfo {author}
  {\bibfnamefont {C.~L.~G.}\ \bibnamefont {Alzar}}, \bibinfo {author} {\bibfnamefont {R.}~\bibnamefont {Geiger}}, \bibinfo {author} {\bibfnamefont {S.}~\bibnamefont {{Guellati-Khelifa}}}, \bibinfo {author} {\bibfnamefont {K.}~\bibnamefont {Hammerer}}, \bibinfo {author} {\bibfnamefont {J.}~\bibnamefont {Harms}}, \bibinfo {author} {\bibfnamefont {J.}~\bibnamefont {Hinderer}}, \bibinfo {author} {\bibfnamefont {M.}~\bibnamefont {Holynski}}, \bibinfo {author} {\bibfnamefont {J.}~\bibnamefont {Junca}}, \bibinfo {author} {\bibfnamefont {S.}~\bibnamefont {Katsanevas}}, \bibinfo {author} {\bibfnamefont {C.}~\bibnamefont {Klempt}}, \bibinfo {author} {\bibfnamefont {C.}~\bibnamefont {Kozanitis}}, \bibinfo {author} {\bibfnamefont {M.}~\bibnamefont {Krutzik}}, \bibinfo {author} {\bibfnamefont {A.}~\bibnamefont {Landragin}}, \bibinfo {author} {\bibfnamefont {I.~L.}\ \bibnamefont {Roche}}, \bibinfo {author} {\bibfnamefont {B.}~\bibnamefont {Leykauf}}, \bibinfo {author} {\bibfnamefont {Y.-H.}\ \bibnamefont {Lien}}, \bibinfo
  {author} {\bibfnamefont {S.}~\bibnamefont {Loriani}}, \bibinfo {author} {\bibfnamefont {S.}~\bibnamefont {Merlet}}, \bibinfo {author} {\bibfnamefont {M.}~\bibnamefont {Merzougui}}, \bibinfo {author} {\bibfnamefont {M.}~\bibnamefont {Nofrarias}}, \bibinfo {author} {\bibfnamefont {P.}~\bibnamefont {Papadakos}}, \bibinfo {author} {\bibfnamefont {F.~P.}\ \bibnamefont {dos Santos}}, \bibinfo {author} {\bibfnamefont {A.}~\bibnamefont {Peters}}, \bibinfo {author} {\bibfnamefont {D.}~\bibnamefont {Plexousakis}}, \bibinfo {author} {\bibfnamefont {M.}~\bibnamefont {Prevedelli}}, \bibinfo {author} {\bibfnamefont {E.~M.}\ \bibnamefont {Rasel}}, \bibinfo {author} {\bibfnamefont {Y.}~\bibnamefont {Rogister}}, \bibinfo {author} {\bibfnamefont {S.}~\bibnamefont {Rosat}}, \bibinfo {author} {\bibfnamefont {A.}~\bibnamefont {Roura}}, \bibinfo {author} {\bibfnamefont {D.~O.}\ \bibnamefont {Sabulsky}}, \bibinfo {author} {\bibfnamefont {V.}~\bibnamefont {Schkolnik}}, \bibinfo {author} {\bibfnamefont {D.}~\bibnamefont
  {Schlippert}}, \bibinfo {author} {\bibfnamefont {C.}~\bibnamefont {Schubert}}, \bibinfo {author} {\bibfnamefont {L.}~\bibnamefont {Sidorenkov}}, \bibinfo {author} {\bibfnamefont {J.-N.}\ \bibnamefont {Siem{\ss}}}, \bibinfo {author} {\bibfnamefont {C.~F.}\ \bibnamefont {Sopuerta}}, \bibinfo {author} {\bibfnamefont {F.}~\bibnamefont {Sorrentino}}, \bibinfo {author} {\bibfnamefont {C.}~\bibnamefont {Struckmann}}, \bibinfo {author} {\bibfnamefont {G.~M.}\ \bibnamefont {Tino}}, \bibinfo {author} {\bibfnamefont {G.}~\bibnamefont {Tsagkatakis}}, \bibinfo {author} {\bibfnamefont {A.}~\bibnamefont {Vicer{\'e}}}, \bibinfo {author} {\bibfnamefont {W.}~\bibnamefont {von Klitzing}}, \bibinfo {author} {\bibfnamefont {L.}~\bibnamefont {Woerner}},\ and\ \bibinfo {author} {\bibfnamefont {X.}~\bibnamefont {Zou}},\ }\href {https://doi.org/10.1088/1361-6382/aba80e} {\bibfield  {journal} {\bibinfo  {journal} {Classical and Quantum Gravity}\ }\textbf {\bibinfo {volume} {37}},\ \bibinfo {pages} {225017} (\bibinfo {year}
  {2020})}\BibitemShut {NoStop}%
\bibitem [{\citenamefont {Abe}\ \emph {et~al.}(2021)\citenamefont {Abe}, \citenamefont {Adamson}, \citenamefont {Borcean}, \citenamefont {Bortoletto}, \citenamefont {Bridges}, \citenamefont {Carman}, \citenamefont {Chattopadhyay}, \citenamefont {Coleman}, \citenamefont {Curfman}, \citenamefont {DeRose}, \citenamefont {Deshpande}, \citenamefont {Dimopoulos}, \citenamefont {Foot}, \citenamefont {Frisch}, \citenamefont {Garber}, \citenamefont {Geer}, \citenamefont {Gibson}, \citenamefont {Glick}, \citenamefont {Graham}, \citenamefont {Hahn}, \citenamefont {Harnik}, \citenamefont {Hawkins}, \citenamefont {Hindley}, \citenamefont {Hogan}, \citenamefont {Jiang}, \citenamefont {Kasevich}, \citenamefont {Kellett}, \citenamefont {Kiburg}, \citenamefont {Kovachy}, \citenamefont {Lykken}, \citenamefont {{March-Russell}}, \citenamefont {Mitchell}, \citenamefont {Murphy}, \citenamefont {Nantel}, \citenamefont {Nobrega}, \citenamefont {Plunkett}, \citenamefont {Rajendran}, \citenamefont {Rudolph}, \citenamefont {Sachdeva},
  \citenamefont {Safdari}, \citenamefont {Santucci}, \citenamefont {Schwartzman}, \citenamefont {Shipsey}, \citenamefont {Swan}, \citenamefont {Valerio}, \citenamefont {Vasonis}, \citenamefont {Wang},\ and\ \citenamefont {Wilkason}}]{Abe2021}%
  \BibitemOpen
  \bibfield  {author} {\bibinfo {author} {\bibfnamefont {M.}~\bibnamefont {Abe}}, \bibinfo {author} {\bibfnamefont {P.}~\bibnamefont {Adamson}}, \bibinfo {author} {\bibfnamefont {M.}~\bibnamefont {Borcean}}, \bibinfo {author} {\bibfnamefont {D.}~\bibnamefont {Bortoletto}}, \bibinfo {author} {\bibfnamefont {K.}~\bibnamefont {Bridges}}, \bibinfo {author} {\bibfnamefont {S.~P.}\ \bibnamefont {Carman}}, \bibinfo {author} {\bibfnamefont {S.}~\bibnamefont {Chattopadhyay}}, \bibinfo {author} {\bibfnamefont {J.}~\bibnamefont {Coleman}}, \bibinfo {author} {\bibfnamefont {N.~M.}\ \bibnamefont {Curfman}}, \bibinfo {author} {\bibfnamefont {K.}~\bibnamefont {DeRose}}, \bibinfo {author} {\bibfnamefont {T.}~\bibnamefont {Deshpande}}, \bibinfo {author} {\bibfnamefont {S.}~\bibnamefont {Dimopoulos}}, \bibinfo {author} {\bibfnamefont {C.~J.}\ \bibnamefont {Foot}}, \bibinfo {author} {\bibfnamefont {J.~C.}\ \bibnamefont {Frisch}}, \bibinfo {author} {\bibfnamefont {B.~E.}\ \bibnamefont {Garber}}, \bibinfo {author} {\bibfnamefont
  {S.}~\bibnamefont {Geer}}, \bibinfo {author} {\bibfnamefont {V.}~\bibnamefont {Gibson}}, \bibinfo {author} {\bibfnamefont {J.}~\bibnamefont {Glick}}, \bibinfo {author} {\bibfnamefont {P.~W.}\ \bibnamefont {Graham}}, \bibinfo {author} {\bibfnamefont {S.~R.}\ \bibnamefont {Hahn}}, \bibinfo {author} {\bibfnamefont {R.}~\bibnamefont {Harnik}}, \bibinfo {author} {\bibfnamefont {L.}~\bibnamefont {Hawkins}}, \bibinfo {author} {\bibfnamefont {S.}~\bibnamefont {Hindley}}, \bibinfo {author} {\bibfnamefont {J.~M.}\ \bibnamefont {Hogan}}, \bibinfo {author} {\bibfnamefont {Y.}~\bibnamefont {Jiang}}, \bibinfo {author} {\bibfnamefont {M.~A.}\ \bibnamefont {Kasevich}}, \bibinfo {author} {\bibfnamefont {R.~J.}\ \bibnamefont {Kellett}}, \bibinfo {author} {\bibfnamefont {M.}~\bibnamefont {Kiburg}}, \bibinfo {author} {\bibfnamefont {T.}~\bibnamefont {Kovachy}}, \bibinfo {author} {\bibfnamefont {J.~D.}\ \bibnamefont {Lykken}}, \bibinfo {author} {\bibfnamefont {J.}~\bibnamefont {{March-Russell}}}, \bibinfo {author}
  {\bibfnamefont {J.}~\bibnamefont {Mitchell}}, \bibinfo {author} {\bibfnamefont {M.}~\bibnamefont {Murphy}}, \bibinfo {author} {\bibfnamefont {M.}~\bibnamefont {Nantel}}, \bibinfo {author} {\bibfnamefont {L.~E.}\ \bibnamefont {Nobrega}}, \bibinfo {author} {\bibfnamefont {R.~K.}\ \bibnamefont {Plunkett}}, \bibinfo {author} {\bibfnamefont {S.}~\bibnamefont {Rajendran}}, \bibinfo {author} {\bibfnamefont {J.}~\bibnamefont {Rudolph}}, \bibinfo {author} {\bibfnamefont {N.}~\bibnamefont {Sachdeva}}, \bibinfo {author} {\bibfnamefont {M.}~\bibnamefont {Safdari}}, \bibinfo {author} {\bibfnamefont {J.~K.}\ \bibnamefont {Santucci}}, \bibinfo {author} {\bibfnamefont {A.~G.}\ \bibnamefont {Schwartzman}}, \bibinfo {author} {\bibfnamefont {I.}~\bibnamefont {Shipsey}}, \bibinfo {author} {\bibfnamefont {H.}~\bibnamefont {Swan}}, \bibinfo {author} {\bibfnamefont {L.~R.}\ \bibnamefont {Valerio}}, \bibinfo {author} {\bibfnamefont {A.}~\bibnamefont {Vasonis}}, \bibinfo {author} {\bibfnamefont {Y.}~\bibnamefont {Wang}},\ and\
  \bibinfo {author} {\bibfnamefont {T.}~\bibnamefont {Wilkason}},\ }\href {https://doi.org/10.1088/2058-9565/abf719} {\bibfield  {journal} {\bibinfo  {journal} {Quantum Science and Technology}\ }\textbf {\bibinfo {volume} {6}},\ \bibinfo {pages} {044003} (\bibinfo {year} {2021})}\BibitemShut {NoStop}%
\bibitem [{\citenamefont {Cheiney}\ \emph {et~al.}(2018)\citenamefont {Cheiney}, \citenamefont {Fouch{\'e}}, \citenamefont {Templier}, \citenamefont {Napolitano}, \citenamefont {Battelier}, \citenamefont {Bouyer},\ and\ \citenamefont {Barrett}}]{Cheiney2018}%
  \BibitemOpen
  \bibfield  {author} {\bibinfo {author} {\bibfnamefont {P.}~\bibnamefont {Cheiney}}, \bibinfo {author} {\bibfnamefont {L.}~\bibnamefont {Fouch{\'e}}}, \bibinfo {author} {\bibfnamefont {S.}~\bibnamefont {Templier}}, \bibinfo {author} {\bibfnamefont {F.}~\bibnamefont {Napolitano}}, \bibinfo {author} {\bibfnamefont {B.}~\bibnamefont {Battelier}}, \bibinfo {author} {\bibfnamefont {P.}~\bibnamefont {Bouyer}},\ and\ \bibinfo {author} {\bibfnamefont {B.}~\bibnamefont {Barrett}},\ }\href {https://doi.org/10.1103/PhysRevApplied.10.034030} {\bibfield  {journal} {\bibinfo  {journal} {Physical Review Applied}\ }\textbf {\bibinfo {volume} {10}},\ \bibinfo {pages} {034030} (\bibinfo {year} {2018})}\BibitemShut {NoStop}%
\bibitem [{\citenamefont {Geiger}\ \emph {et~al.}(2020)\citenamefont {Geiger}, \citenamefont {Landragin}, \citenamefont {Merlet},\ and\ \citenamefont {Pereira Dos~Santos}}]{Geiger2020}%
  \BibitemOpen
  \bibfield  {author} {\bibinfo {author} {\bibfnamefont {R.}~\bibnamefont {Geiger}}, \bibinfo {author} {\bibfnamefont {A.}~\bibnamefont {Landragin}}, \bibinfo {author} {\bibfnamefont {S.}~\bibnamefont {Merlet}},\ and\ \bibinfo {author} {\bibfnamefont {F.}~\bibnamefont {Pereira Dos~Santos}},\ }\href {https://doi.org/10.1116/5.0009093} {\bibfield  {journal} {\bibinfo  {journal} {AVS Quantum Science}\ }\textbf {\bibinfo {volume} {2}},\ \bibinfo {pages} {024702} (\bibinfo {year} {2020})}\BibitemShut {NoStop}%
\bibitem [{\citenamefont {Hensel}\ \emph {et~al.}(2021)\citenamefont {Hensel}, \citenamefont {Loriani}, \citenamefont {Schubert}, \citenamefont {Fitzek}, \citenamefont {Abend}, \citenamefont {Ahlers}, \citenamefont {Siem{\ss}}, \citenamefont {Hammerer}, \citenamefont {Rasel},\ and\ \citenamefont {Gaaloul}}]{Hensel2021}%
  \BibitemOpen
  \bibfield  {author} {\bibinfo {author} {\bibfnamefont {T.}~\bibnamefont {Hensel}}, \bibinfo {author} {\bibfnamefont {S.}~\bibnamefont {Loriani}}, \bibinfo {author} {\bibfnamefont {C.}~\bibnamefont {Schubert}}, \bibinfo {author} {\bibfnamefont {F.}~\bibnamefont {Fitzek}}, \bibinfo {author} {\bibfnamefont {S.}~\bibnamefont {Abend}}, \bibinfo {author} {\bibfnamefont {H.}~\bibnamefont {Ahlers}}, \bibinfo {author} {\bibfnamefont {J.-N.}\ \bibnamefont {Siem{\ss}}}, \bibinfo {author} {\bibfnamefont {K.}~\bibnamefont {Hammerer}}, \bibinfo {author} {\bibfnamefont {E.~M.}\ \bibnamefont {Rasel}},\ and\ \bibinfo {author} {\bibfnamefont {N.}~\bibnamefont {Gaaloul}},\ }\href {https://doi.org/10.1140/epjd/s10053-021-00069-9} {\bibfield  {journal} {\bibinfo  {journal} {The European Physical Journal D}\ }\textbf {\bibinfo {volume} {75}},\ \bibinfo {pages} {108} (\bibinfo {year} {2021})}\BibitemShut {NoStop}%
\bibitem [{\citenamefont {Struckmann}\ \emph {et~al.}(2024)\citenamefont {Struckmann}, \citenamefont {Corgier}, \citenamefont {Loriani}, \citenamefont {Kleinsteinberg}, \citenamefont {Gox}, \citenamefont {Giese}, \citenamefont {M{\'e}tris}, \citenamefont {Gaaloul},\ and\ \citenamefont {Wolf}}]{Struckmann2024}%
  \BibitemOpen
  \bibfield  {author} {\bibinfo {author} {\bibfnamefont {C.}~\bibnamefont {Struckmann}}, \bibinfo {author} {\bibfnamefont {R.}~\bibnamefont {Corgier}}, \bibinfo {author} {\bibfnamefont {S.}~\bibnamefont {Loriani}}, \bibinfo {author} {\bibfnamefont {G.}~\bibnamefont {Kleinsteinberg}}, \bibinfo {author} {\bibfnamefont {N.}~\bibnamefont {Gox}}, \bibinfo {author} {\bibfnamefont {E.}~\bibnamefont {Giese}}, \bibinfo {author} {\bibfnamefont {G.}~\bibnamefont {M{\'e}tris}}, \bibinfo {author} {\bibfnamefont {N.}~\bibnamefont {Gaaloul}},\ and\ \bibinfo {author} {\bibfnamefont {P.}~\bibnamefont {Wolf}},\ }\href {https://doi.org/10.1103/PhysRevD.109.064010} {\bibfield  {journal} {\bibinfo  {journal} {Physical Review D}\ }\textbf {\bibinfo {volume} {109}},\ \bibinfo {pages} {064010} (\bibinfo {year} {2024})}\BibitemShut {NoStop}%
\bibitem [{\citenamefont {Qi}\ \emph {et~al.}(2021)\citenamefont {Qi}, \citenamefont {Chiaverini}, \citenamefont {Espin{\'o}s}, \citenamefont {Palmero},\ and\ \citenamefont {Muga}}]{Qi2021}%
  \BibitemOpen
  \bibfield  {author} {\bibinfo {author} {\bibfnamefont {L.}~\bibnamefont {Qi}}, \bibinfo {author} {\bibfnamefont {J.}~\bibnamefont {Chiaverini}}, \bibinfo {author} {\bibfnamefont {H.}~\bibnamefont {Espin{\'o}s}}, \bibinfo {author} {\bibfnamefont {M.}~\bibnamefont {Palmero}},\ and\ \bibinfo {author} {\bibfnamefont {J.~G.}\ \bibnamefont {Muga}},\ }\href {https://doi.org/10.1209/0295-5075/134/23001} {\bibfield  {journal} {\bibinfo  {journal} {Europhysics Letters}\ }\textbf {\bibinfo {volume} {134}},\ \bibinfo {pages} {23001} (\bibinfo {year} {2021})}\BibitemShut {NoStop}%
\bibitem [{\citenamefont {Gaaloul}\ \emph {et~al.}(2022)\citenamefont {Gaaloul}, \citenamefont {Meister}, \citenamefont {Corgier}, \citenamefont {Pichery}, \citenamefont {Boegel}, \citenamefont {Herr}, \citenamefont {Ahlers}, \citenamefont {Charron}, \citenamefont {Williams}, \citenamefont {Thompson}, \citenamefont {Schleich}, \citenamefont {Rasel},\ and\ \citenamefont {Bigelow}}]{Gaaloul2022}%
  \BibitemOpen
  \bibfield  {author} {\bibinfo {author} {\bibfnamefont {N.}~\bibnamefont {Gaaloul}}, \bibinfo {author} {\bibfnamefont {M.}~\bibnamefont {Meister}}, \bibinfo {author} {\bibfnamefont {R.}~\bibnamefont {Corgier}}, \bibinfo {author} {\bibfnamefont {A.}~\bibnamefont {Pichery}}, \bibinfo {author} {\bibfnamefont {P.}~\bibnamefont {Boegel}}, \bibinfo {author} {\bibfnamefont {W.}~\bibnamefont {Herr}}, \bibinfo {author} {\bibfnamefont {H.}~\bibnamefont {Ahlers}}, \bibinfo {author} {\bibfnamefont {E.}~\bibnamefont {Charron}}, \bibinfo {author} {\bibfnamefont {J.~R.}\ \bibnamefont {Williams}}, \bibinfo {author} {\bibfnamefont {R.~J.}\ \bibnamefont {Thompson}}, \bibinfo {author} {\bibfnamefont {W.~P.}\ \bibnamefont {Schleich}}, \bibinfo {author} {\bibfnamefont {E.~M.}\ \bibnamefont {Rasel}},\ and\ \bibinfo {author} {\bibfnamefont {N.~P.}\ \bibnamefont {Bigelow}},\ }\href {https://doi.org/10.1038/s41467-022-35274-6} {\bibfield  {journal} {\bibinfo  {journal} {Nature Communications}\ }\textbf {\bibinfo {volume} {13}},\
  \bibinfo {pages} {7889} (\bibinfo {year} {2022})}\BibitemShut {NoStop}%
\bibitem [{\citenamefont {Deppner}\ \emph {et~al.}(2021)\citenamefont {Deppner}, \citenamefont {Herr}, \citenamefont {Cornelius}, \citenamefont {Stromberger}, \citenamefont {Sternke}, \citenamefont {Grzeschik}, \citenamefont {Grote}, \citenamefont {Rudolph}, \citenamefont {Herrmann}, \citenamefont {Krutzik}, \citenamefont {Wenzlawski}, \citenamefont {Corgier}, \citenamefont {Charron}, \citenamefont {{Gu{\'e}ry-Odelin}}, \citenamefont {Gaaloul}, \citenamefont {L{\"a}mmerzahl}, \citenamefont {Peters}, \citenamefont {Windpassinger},\ and\ \citenamefont {Rasel}}]{Deppner2021a}%
  \BibitemOpen
  \bibfield  {author} {\bibinfo {author} {\bibfnamefont {C.}~\bibnamefont {Deppner}}, \bibinfo {author} {\bibfnamefont {W.}~\bibnamefont {Herr}}, \bibinfo {author} {\bibfnamefont {M.}~\bibnamefont {Cornelius}}, \bibinfo {author} {\bibfnamefont {P.}~\bibnamefont {Stromberger}}, \bibinfo {author} {\bibfnamefont {T.}~\bibnamefont {Sternke}}, \bibinfo {author} {\bibfnamefont {C.}~\bibnamefont {Grzeschik}}, \bibinfo {author} {\bibfnamefont {A.}~\bibnamefont {Grote}}, \bibinfo {author} {\bibfnamefont {J.}~\bibnamefont {Rudolph}}, \bibinfo {author} {\bibfnamefont {S.}~\bibnamefont {Herrmann}}, \bibinfo {author} {\bibfnamefont {M.}~\bibnamefont {Krutzik}}, \bibinfo {author} {\bibfnamefont {A.}~\bibnamefont {Wenzlawski}}, \bibinfo {author} {\bibfnamefont {R.}~\bibnamefont {Corgier}}, \bibinfo {author} {\bibfnamefont {E.}~\bibnamefont {Charron}}, \bibinfo {author} {\bibfnamefont {D.}~\bibnamefont {{Gu{\'e}ry-Odelin}}}, \bibinfo {author} {\bibfnamefont {N.}~\bibnamefont {Gaaloul}}, \bibinfo {author} {\bibfnamefont
  {C.}~\bibnamefont {L{\"a}mmerzahl}}, \bibinfo {author} {\bibfnamefont {A.}~\bibnamefont {Peters}}, \bibinfo {author} {\bibfnamefont {P.}~\bibnamefont {Windpassinger}},\ and\ \bibinfo {author} {\bibfnamefont {E.~M.}\ \bibnamefont {Rasel}},\ }\href {https://doi.org/10.1103/PhysRevLett.127.100401} {\bibfield  {journal} {\bibinfo  {journal} {Physical Review Letters}\ }\textbf {\bibinfo {volume} {127}},\ \bibinfo {pages} {100401} (\bibinfo {year} {2021})}\BibitemShut {NoStop}%
\bibitem [{\citenamefont {Peirce}\ \emph {et~al.}(1988)\citenamefont {Peirce}, \citenamefont {Dahleh},\ and\ \citenamefont {Rabitz}}]{Peirce1988}%
  \BibitemOpen
  \bibfield  {author} {\bibinfo {author} {\bibfnamefont {A.~P.}\ \bibnamefont {Peirce}}, \bibinfo {author} {\bibfnamefont {M.~A.}\ \bibnamefont {Dahleh}},\ and\ \bibinfo {author} {\bibfnamefont {H.}~\bibnamefont {Rabitz}},\ }\href {https://doi.org/10.1103/PhysRevA.37.4950} {\bibfield  {journal} {\bibinfo  {journal} {Physical Review A}\ }\textbf {\bibinfo {volume} {37}},\ \bibinfo {pages} {4950} (\bibinfo {year} {1988})}\BibitemShut {NoStop}%
\bibitem [{\citenamefont {J{\"a}ger}\ \emph {et~al.}(2014)\citenamefont {J{\"a}ger}, \citenamefont {Reich}, \citenamefont {Goerz}, \citenamefont {Koch},\ and\ \citenamefont {Hohenester}}]{Jager2014}%
  \BibitemOpen
  \bibfield  {author} {\bibinfo {author} {\bibfnamefont {G.}~\bibnamefont {J{\"a}ger}}, \bibinfo {author} {\bibfnamefont {D.~M.}\ \bibnamefont {Reich}}, \bibinfo {author} {\bibfnamefont {M.~H.}\ \bibnamefont {Goerz}}, \bibinfo {author} {\bibfnamefont {C.~P.}\ \bibnamefont {Koch}},\ and\ \bibinfo {author} {\bibfnamefont {U.}~\bibnamefont {Hohenester}},\ }\href {https://doi.org/10.1103/PhysRevA.90.033628} {\bibfield  {journal} {\bibinfo  {journal} {Physical Review A}\ }\textbf {\bibinfo {volume} {90}},\ \bibinfo {pages} {033628} (\bibinfo {year} {2014})}\BibitemShut {NoStop}%
\bibitem [{\citenamefont {Amri}\ \emph {et~al.}(2019)\citenamefont {Amri}, \citenamefont {Corgier}, \citenamefont {Sugny}, \citenamefont {Rasel}, \citenamefont {Gaaloul},\ and\ \citenamefont {Charron}}]{Amri2019}%
  \BibitemOpen
  \bibfield  {author} {\bibinfo {author} {\bibfnamefont {S.}~\bibnamefont {Amri}}, \bibinfo {author} {\bibfnamefont {R.}~\bibnamefont {Corgier}}, \bibinfo {author} {\bibfnamefont {D.}~\bibnamefont {Sugny}}, \bibinfo {author} {\bibfnamefont {E.~M.}\ \bibnamefont {Rasel}}, \bibinfo {author} {\bibfnamefont {N.}~\bibnamefont {Gaaloul}},\ and\ \bibinfo {author} {\bibfnamefont {E.}~\bibnamefont {Charron}},\ }\href {https://doi.org/10.1038/s41598-019-41784-z} {\bibfield  {journal} {\bibinfo  {journal} {Scientific Reports}\ }\textbf {\bibinfo {volume} {9}},\ \bibinfo {pages} {5346} (\bibinfo {year} {2019})}\BibitemShut {NoStop}%
\bibitem [{\citenamefont {Castin}\ and\ \citenamefont {Dum}(1996)}]{Castin1996}%
  \BibitemOpen
  \bibfield  {author} {\bibinfo {author} {\bibfnamefont {Y.}~\bibnamefont {Castin}}\ and\ \bibinfo {author} {\bibfnamefont {R.}~\bibnamefont {Dum}},\ }\href {https://doi.org/10.1103/PhysRevLett.77.5315} {\bibfield  {journal} {\bibinfo  {journal} {Physical Review Letters}\ }\textbf {\bibinfo {volume} {77}},\ \bibinfo {pages} {5315} (\bibinfo {year} {1996})}\BibitemShut {NoStop}%
\bibitem [{\citenamefont {Elliott}\ \emph {et~al.}(2023)\citenamefont {Elliott}, \citenamefont {Aveline}, \citenamefont {Bigelow}, \citenamefont {Boegel}, \citenamefont {Botsi}, \citenamefont {Charron}, \citenamefont {D'Incao}, \citenamefont {Engels}, \citenamefont {Estrampes}, \citenamefont {Gaaloul}, \citenamefont {Kellogg}, \citenamefont {Kohel}, \citenamefont {Lay}, \citenamefont {Lundblad}, \citenamefont {Meister}, \citenamefont {Mossman}, \citenamefont {M{\"u}ller}, \citenamefont {M{\"u}ller}, \citenamefont {Oudrhiri}, \citenamefont {Phillips}, \citenamefont {Pichery}, \citenamefont {Rasel}, \citenamefont {Sackett}, \citenamefont {Sbroscia}, \citenamefont {Schleich}, \citenamefont {Thompson},\ and\ \citenamefont {Williams}}]{Elliott2023a}%
  \BibitemOpen
  \bibfield  {author} {\bibinfo {author} {\bibfnamefont {E.~R.}\ \bibnamefont {Elliott}}, \bibinfo {author} {\bibfnamefont {D.~C.}\ \bibnamefont {Aveline}}, \bibinfo {author} {\bibfnamefont {N.~P.}\ \bibnamefont {Bigelow}}, \bibinfo {author} {\bibfnamefont {P.}~\bibnamefont {Boegel}}, \bibinfo {author} {\bibfnamefont {S.}~\bibnamefont {Botsi}}, \bibinfo {author} {\bibfnamefont {E.}~\bibnamefont {Charron}}, \bibinfo {author} {\bibfnamefont {J.~P.}\ \bibnamefont {D'Incao}}, \bibinfo {author} {\bibfnamefont {P.}~\bibnamefont {Engels}}, \bibinfo {author} {\bibfnamefont {T.}~\bibnamefont {Estrampes}}, \bibinfo {author} {\bibfnamefont {N.}~\bibnamefont {Gaaloul}}, \bibinfo {author} {\bibfnamefont {J.~R.}\ \bibnamefont {Kellogg}}, \bibinfo {author} {\bibfnamefont {J.~M.}\ \bibnamefont {Kohel}}, \bibinfo {author} {\bibfnamefont {N.~E.}\ \bibnamefont {Lay}}, \bibinfo {author} {\bibfnamefont {N.}~\bibnamefont {Lundblad}}, \bibinfo {author} {\bibfnamefont {M.}~\bibnamefont {Meister}}, \bibinfo {author} {\bibfnamefont
  {M.~E.}\ \bibnamefont {Mossman}}, \bibinfo {author} {\bibfnamefont {G.}~\bibnamefont {M{\"u}ller}}, \bibinfo {author} {\bibfnamefont {H.}~\bibnamefont {M{\"u}ller}}, \bibinfo {author} {\bibfnamefont {K.}~\bibnamefont {Oudrhiri}}, \bibinfo {author} {\bibfnamefont {L.~E.}\ \bibnamefont {Phillips}}, \bibinfo {author} {\bibfnamefont {A.}~\bibnamefont {Pichery}}, \bibinfo {author} {\bibfnamefont {E.~M.}\ \bibnamefont {Rasel}}, \bibinfo {author} {\bibfnamefont {C.~A.}\ \bibnamefont {Sackett}}, \bibinfo {author} {\bibfnamefont {M.}~\bibnamefont {Sbroscia}}, \bibinfo {author} {\bibfnamefont {W.~P.}\ \bibnamefont {Schleich}}, \bibinfo {author} {\bibfnamefont {R.~J.}\ \bibnamefont {Thompson}},\ and\ \bibinfo {author} {\bibfnamefont {J.~R.}\ \bibnamefont {Williams}},\ }\href {https://doi.org/10.1038/s41586-023-06645-w} {\bibfield  {journal} {\bibinfo  {journal} {Nature}\ }\textbf {\bibinfo {volume} {623}},\ \bibinfo {pages} {502} (\bibinfo {year} {2023})}\BibitemShut {NoStop}%
\bibitem [{\citenamefont {Gross}(1963)}]{Gross1963}%
  \BibitemOpen
  \bibfield  {author} {\bibinfo {author} {\bibfnamefont {E.~P.}\ \bibnamefont {Gross}},\ }\href {https://doi.org/10.1063/1.1703944} {\bibfield  {journal} {\bibinfo  {journal} {Journal of Mathematical Physics}\ }\textbf {\bibinfo {volume} {4}},\ \bibinfo {pages} {195} (\bibinfo {year} {1963})}\BibitemShut {NoStop}%
\bibitem [{\citenamefont {Pitaevskii}(1961)}]{Pitaevskii1961a}%
  \BibitemOpen
  \bibfield  {author} {\bibinfo {author} {\bibfnamefont {L.~P.}\ \bibnamefont {Pitaevskii}},\ }\href@noop {} {\bibfield  {journal} {\bibinfo  {journal} {Sov. Phys. JETP}\ }\textbf {\bibinfo {volume} {13}},\ \bibinfo {pages} {451} (\bibinfo {year} {1961})}\BibitemShut {NoStop}%
\bibitem [{\citenamefont {Pichery}\ \emph {et~al.}(2023)\citenamefont {Pichery}, \citenamefont {Meister}, \citenamefont {Piest}, \citenamefont {B{\"o}hm}, \citenamefont {Rasel}, \citenamefont {Charron},\ and\ \citenamefont {Gaaloul}}]{Pichery2023}%
  \BibitemOpen
  \bibfield  {author} {\bibinfo {author} {\bibfnamefont {A.}~\bibnamefont {Pichery}}, \bibinfo {author} {\bibfnamefont {M.}~\bibnamefont {Meister}}, \bibinfo {author} {\bibfnamefont {B.}~\bibnamefont {Piest}}, \bibinfo {author} {\bibfnamefont {J.}~\bibnamefont {B{\"o}hm}}, \bibinfo {author} {\bibfnamefont {E.~M.}\ \bibnamefont {Rasel}}, \bibinfo {author} {\bibfnamefont {E.}~\bibnamefont {Charron}},\ and\ \bibinfo {author} {\bibfnamefont {N.}~\bibnamefont {Gaaloul}},\ }\href {https://doi.org/10.1116/5.0163850} {\bibfield  {journal} {\bibinfo  {journal} {AVS Quantum Science}\ }\textbf {\bibinfo {volume} {5}},\ \bibinfo {pages} {044401} (\bibinfo {year} {2023})}\BibitemShut {NoStop}%
\bibitem [{\citenamefont {Shahriari}\ \emph {et~al.}(2016)\citenamefont {Shahriari}, \citenamefont {Swersky}, \citenamefont {Wang}, \citenamefont {Adams},\ and\ \citenamefont {{de Freitas}}}]{Shahriari2016}%
  \BibitemOpen
  \bibfield  {author} {\bibinfo {author} {\bibfnamefont {B.}~\bibnamefont {Shahriari}}, \bibinfo {author} {\bibfnamefont {K.}~\bibnamefont {Swersky}}, \bibinfo {author} {\bibfnamefont {Z.}~\bibnamefont {Wang}}, \bibinfo {author} {\bibfnamefont {R.~P.}\ \bibnamefont {Adams}},\ and\ \bibinfo {author} {\bibfnamefont {N.}~\bibnamefont {{de Freitas}}},\ }\href {https://doi.org/10.1109/JPROC.2015.2494218} {\bibfield  {journal} {\bibinfo  {journal} {Proceedings of the IEEE}\ }\textbf {\bibinfo {volume} {104}},\ \bibinfo {pages} {148} (\bibinfo {year} {2016})}\BibitemShut {NoStop}%
\bibitem [{\citenamefont {Rasmussen}\ and\ \citenamefont {Williams}(2006)}]{Rasmussen2006}%
  \BibitemOpen
  \bibfield  {author} {\bibinfo {author} {\bibfnamefont {C.~E.}\ \bibnamefont {Rasmussen}}\ and\ \bibinfo {author} {\bibfnamefont {C.~K.~I.}\ \bibnamefont {Williams}},\ }\href@noop {} {\emph {\bibinfo {title} {Gaussian {{Processes}} for {{Machine Learning}}}}},\ \bibinfo {edition} {3rd}\ ed.,\ Vol.~\bibinfo {volume} {2}\ (\bibinfo  {publisher} {Cambridge, MA: MIT press},\ \bibinfo {year} {2006})\BibitemShut {NoStop}%
\bibitem [{\citenamefont {Schneider}\ \emph {et~al.}(2019)\citenamefont {Schneider}, \citenamefont {Garcia~Santiago}, \citenamefont {Soltwisch}, \citenamefont {Hammerschmidt}, \citenamefont {Burger},\ and\ \citenamefont {Rockstuhl}}]{Schneider2019}%
  \BibitemOpen
  \bibfield  {author} {\bibinfo {author} {\bibfnamefont {P.-I.}\ \bibnamefont {Schneider}}, \bibinfo {author} {\bibfnamefont {X.}~\bibnamefont {Garcia~Santiago}}, \bibinfo {author} {\bibfnamefont {V.}~\bibnamefont {Soltwisch}}, \bibinfo {author} {\bibfnamefont {M.}~\bibnamefont {Hammerschmidt}}, \bibinfo {author} {\bibfnamefont {S.}~\bibnamefont {Burger}},\ and\ \bibinfo {author} {\bibfnamefont {C.}~\bibnamefont {Rockstuhl}},\ }\href {https://doi.org/10.1021/acsphotonics.9b00706} {\bibfield  {journal} {\bibinfo  {journal} {ACS Photonics}\ }\textbf {\bibinfo {volume} {6}},\ \bibinfo {pages} {2726} (\bibinfo {year} {2019})}\BibitemShut {NoStop}%
\bibitem [{\citenamefont {Plock}\ \emph {et~al.}(2022)\citenamefont {Plock}, \citenamefont {Andrle}, \citenamefont {Burger},\ and\ \citenamefont {Schneider}}]{Plock2022}%
  \BibitemOpen
  \bibfield  {author} {\bibinfo {author} {\bibfnamefont {M.}~\bibnamefont {Plock}}, \bibinfo {author} {\bibfnamefont {K.}~\bibnamefont {Andrle}}, \bibinfo {author} {\bibfnamefont {S.}~\bibnamefont {Burger}},\ and\ \bibinfo {author} {\bibfnamefont {P.-I.}\ \bibnamefont {Schneider}},\ }\href {https://doi.org/10.1002/adts.202200112} {\bibfield  {journal} {\bibinfo  {journal} {Advanced Theory and Simulations}\ }\textbf {\bibinfo {volume} {5}},\ \bibinfo {pages} {2200112} (\bibinfo {year} {2022})}\BibitemShut {NoStop}%
\bibitem [{\citenamefont {Anton}\ \emph {et~al.}(2024)\citenamefont {Anton}, \citenamefont {Henderson}, \citenamefont {Da~Ros}, \citenamefont {Sekulic}, \citenamefont {Burger}, \citenamefont {Schneider},\ and\ \citenamefont {Krutzik}}]{Anton2024}%
  \BibitemOpen
  \bibfield  {author} {\bibinfo {author} {\bibfnamefont {O.}~\bibnamefont {Anton}}, \bibinfo {author} {\bibfnamefont {V.~A.}\ \bibnamefont {Henderson}}, \bibinfo {author} {\bibfnamefont {E.}~\bibnamefont {Da~Ros}}, \bibinfo {author} {\bibfnamefont {I.}~\bibnamefont {Sekulic}}, \bibinfo {author} {\bibfnamefont {S.}~\bibnamefont {Burger}}, \bibinfo {author} {\bibfnamefont {P.-I.}\ \bibnamefont {Schneider}},\ and\ \bibinfo {author} {\bibfnamefont {M.}~\bibnamefont {Krutzik}},\ }\bibfield  {journal} {\bibinfo  {journal} {Machine Learning: Science and Technology}\ }\href {https://doi.org/10.1088/2632-2153/ad3cb6} {10.1088/2632-2153/ad3cb6} (\bibinfo {year} {2024})\BibitemShut {NoStop}%
\bibitem [{\citenamefont {{de Boor}}(1978)}]{deBoor1978}%
  \BibitemOpen
  \bibfield  {author} {\bibinfo {author} {\bibfnamefont {C.}~\bibnamefont {{de Boor}}},\ }\href@noop {} {\bibfield  {journal} {\bibinfo  {journal} {New York: springer-verlag}\ }\textbf {\bibinfo {volume} {27}} (\bibinfo {year} {1978})}\BibitemShut {NoStop}%
\bibitem [{\citenamefont {Corgier}\ \emph {et~al.}(2018)\citenamefont {Corgier}, \citenamefont {Amri}, \citenamefont {Herr}, \citenamefont {Ahlers}, \citenamefont {Rudolph}, \citenamefont {{Gu{\'e}ry-Odelin}}, \citenamefont {Rasel}, \citenamefont {Charron},\ and\ \citenamefont {Gaaloul}}]{Corgier2018}%
  \BibitemOpen
  \bibfield  {author} {\bibinfo {author} {\bibfnamefont {R.}~\bibnamefont {Corgier}}, \bibinfo {author} {\bibfnamefont {S.}~\bibnamefont {Amri}}, \bibinfo {author} {\bibfnamefont {W.}~\bibnamefont {Herr}}, \bibinfo {author} {\bibfnamefont {H.}~\bibnamefont {Ahlers}}, \bibinfo {author} {\bibfnamefont {J.}~\bibnamefont {Rudolph}}, \bibinfo {author} {\bibfnamefont {D.}~\bibnamefont {{Gu{\'e}ry-Odelin}}}, \bibinfo {author} {\bibfnamefont {E.~M.}\ \bibnamefont {Rasel}}, \bibinfo {author} {\bibfnamefont {E.}~\bibnamefont {Charron}},\ and\ \bibinfo {author} {\bibfnamefont {N.}~\bibnamefont {Gaaloul}},\ }\href {https://doi.org/10.1088/1367-2630/aabdfc} {\bibfield  {journal} {\bibinfo  {journal} {New Journal of Physics}\ }\textbf {\bibinfo {volume} {20}},\ \bibinfo {pages} {055002} (\bibinfo {year} {2018})}\BibitemShut {NoStop}%
\bibitem [{\citenamefont {Becker}\ \emph {et~al.}(2018)\citenamefont {Becker}, \citenamefont {Lachmann}, \citenamefont {Seidel}, \citenamefont {Ahlers}, \citenamefont {Dinkelaker}, \citenamefont {Grosse}, \citenamefont {Hellmig}, \citenamefont {M{\"u}ntinga}, \citenamefont {Schkolnik}, \citenamefont {Wendrich}, \citenamefont {Wenzlawski}, \citenamefont {Weps}, \citenamefont {Corgier}, \citenamefont {Franz}, \citenamefont {Gaaloul}, \citenamefont {Herr}, \citenamefont {L{\"u}dtke}, \citenamefont {Popp}, \citenamefont {Amri}, \citenamefont {Duncker}, \citenamefont {Erbe}, \citenamefont {Kohfeldt}, \citenamefont {{Kubelka-Lange}}, \citenamefont {Braxmaier}, \citenamefont {Charron}, \citenamefont {Ertmer}, \citenamefont {Krutzik}, \citenamefont {L{\"a}mmerzahl}, \citenamefont {Peters}, \citenamefont {Schleich}, \citenamefont {Sengstock}, \citenamefont {Walser}, \citenamefont {Wicht}, \citenamefont {Windpassinger},\ and\ \citenamefont {Rasel}}]{Becker2018}%
  \BibitemOpen
  \bibfield  {author} {\bibinfo {author} {\bibfnamefont {D.}~\bibnamefont {Becker}}, \bibinfo {author} {\bibfnamefont {M.~D.}\ \bibnamefont {Lachmann}}, \bibinfo {author} {\bibfnamefont {S.~T.}\ \bibnamefont {Seidel}}, \bibinfo {author} {\bibfnamefont {H.}~\bibnamefont {Ahlers}}, \bibinfo {author} {\bibfnamefont {A.~N.}\ \bibnamefont {Dinkelaker}}, \bibinfo {author} {\bibfnamefont {J.}~\bibnamefont {Grosse}}, \bibinfo {author} {\bibfnamefont {O.}~\bibnamefont {Hellmig}}, \bibinfo {author} {\bibfnamefont {H.}~\bibnamefont {M{\"u}ntinga}}, \bibinfo {author} {\bibfnamefont {V.}~\bibnamefont {Schkolnik}}, \bibinfo {author} {\bibfnamefont {T.}~\bibnamefont {Wendrich}}, \bibinfo {author} {\bibfnamefont {A.}~\bibnamefont {Wenzlawski}}, \bibinfo {author} {\bibfnamefont {B.}~\bibnamefont {Weps}}, \bibinfo {author} {\bibfnamefont {R.}~\bibnamefont {Corgier}}, \bibinfo {author} {\bibfnamefont {T.}~\bibnamefont {Franz}}, \bibinfo {author} {\bibfnamefont {N.}~\bibnamefont {Gaaloul}}, \bibinfo {author} {\bibfnamefont
  {W.}~\bibnamefont {Herr}}, \bibinfo {author} {\bibfnamefont {D.}~\bibnamefont {L{\"u}dtke}}, \bibinfo {author} {\bibfnamefont {M.}~\bibnamefont {Popp}}, \bibinfo {author} {\bibfnamefont {S.}~\bibnamefont {Amri}}, \bibinfo {author} {\bibfnamefont {H.}~\bibnamefont {Duncker}}, \bibinfo {author} {\bibfnamefont {M.}~\bibnamefont {Erbe}}, \bibinfo {author} {\bibfnamefont {A.}~\bibnamefont {Kohfeldt}}, \bibinfo {author} {\bibfnamefont {A.}~\bibnamefont {{Kubelka-Lange}}}, \bibinfo {author} {\bibfnamefont {C.}~\bibnamefont {Braxmaier}}, \bibinfo {author} {\bibfnamefont {E.}~\bibnamefont {Charron}}, \bibinfo {author} {\bibfnamefont {W.}~\bibnamefont {Ertmer}}, \bibinfo {author} {\bibfnamefont {M.}~\bibnamefont {Krutzik}}, \bibinfo {author} {\bibfnamefont {C.}~\bibnamefont {L{\"a}mmerzahl}}, \bibinfo {author} {\bibfnamefont {A.}~\bibnamefont {Peters}}, \bibinfo {author} {\bibfnamefont {W.~P.}\ \bibnamefont {Schleich}}, \bibinfo {author} {\bibfnamefont {K.}~\bibnamefont {Sengstock}}, \bibinfo {author} {\bibfnamefont
  {R.}~\bibnamefont {Walser}}, \bibinfo {author} {\bibfnamefont {A.}~\bibnamefont {Wicht}}, \bibinfo {author} {\bibfnamefont {P.}~\bibnamefont {Windpassinger}},\ and\ \bibinfo {author} {\bibfnamefont {E.~M.}\ \bibnamefont {Rasel}},\ }\href {https://doi.org/10.1038/s41586-018-0605-1} {\bibfield  {journal} {\bibinfo  {journal} {Nature}\ }\textbf {\bibinfo {volume} {562}},\ \bibinfo {pages} {391} (\bibinfo {year} {2018})}\BibitemShut {NoStop}%
\bibitem [{\citenamefont {Weber}\ \emph {et~al.}(2003)\citenamefont {Weber}, \citenamefont {Herbig}, \citenamefont {Mark}, \citenamefont {N{\"a}gerl},\ and\ \citenamefont {Grimm}}]{Weber2003}%
  \BibitemOpen
  \bibfield  {author} {\bibinfo {author} {\bibfnamefont {T.}~\bibnamefont {Weber}}, \bibinfo {author} {\bibfnamefont {J.}~\bibnamefont {Herbig}}, \bibinfo {author} {\bibfnamefont {M.}~\bibnamefont {Mark}}, \bibinfo {author} {\bibfnamefont {H.-C.}\ \bibnamefont {N{\"a}gerl}},\ and\ \bibinfo {author} {\bibfnamefont {R.}~\bibnamefont {Grimm}},\ }\href {https://doi.org/10.1126/science.1079699} {\bibfield  {journal} {\bibinfo  {journal} {Science}\ }\textbf {\bibinfo {volume} {299}},\ \bibinfo {pages} {232} (\bibinfo {year} {2003})}\BibitemShut {NoStop}%
\bibitem [{\citenamefont {Herbst}\ \emph {et~al.}(2024)\citenamefont {Herbst}, \citenamefont {Estrampes}, \citenamefont {Albers}, \citenamefont {Corgier}, \citenamefont {Stolzenberg}, \citenamefont {Bode}, \citenamefont {Charron}, \citenamefont {Rasel}, \citenamefont {Gaaloul},\ and\ \citenamefont {Schlippert}}]{Herbst2024a}%
  \BibitemOpen
  \bibfield  {author} {\bibinfo {author} {\bibfnamefont {A.}~\bibnamefont {Herbst}}, \bibinfo {author} {\bibfnamefont {T.}~\bibnamefont {Estrampes}}, \bibinfo {author} {\bibfnamefont {H.}~\bibnamefont {Albers}}, \bibinfo {author} {\bibfnamefont {R.}~\bibnamefont {Corgier}}, \bibinfo {author} {\bibfnamefont {K.}~\bibnamefont {Stolzenberg}}, \bibinfo {author} {\bibfnamefont {S.}~\bibnamefont {Bode}}, \bibinfo {author} {\bibfnamefont {E.}~\bibnamefont {Charron}}, \bibinfo {author} {\bibfnamefont {E.~M.}\ \bibnamefont {Rasel}}, \bibinfo {author} {\bibfnamefont {N.}~\bibnamefont {Gaaloul}},\ and\ \bibinfo {author} {\bibfnamefont {D.}~\bibnamefont {Schlippert}},\ }\href {https://doi.org/10.1038/s42005-024-01621-w} {\bibfield  {journal} {\bibinfo  {journal} {Communications Physics}\ }\textbf {\bibinfo {volume} {7}},\ \bibinfo {pages} {1} (\bibinfo {year} {2024})}\BibitemShut {NoStop}%
\bibitem [{\citenamefont {Masuda}\ and\ \citenamefont {Nakamura}(2009)}]{Masuda2009}%
  \BibitemOpen
  \bibfield  {author} {\bibinfo {author} {\bibfnamefont {S.}~\bibnamefont {Masuda}}\ and\ \bibinfo {author} {\bibfnamefont {K.}~\bibnamefont {Nakamura}},\ }\href {https://doi.org/10.1098/rspa.2009.0446} {\bibfield  {journal} {\bibinfo  {journal} {Proceedings of the Royal Society A: Mathematical, Physical and Engineering Sciences}\ }\textbf {\bibinfo {volume} {466}},\ \bibinfo {pages} {1135} (\bibinfo {year} {2009})}\BibitemShut {NoStop}%
\bibitem [{\citenamefont {Torrontegui}\ \emph {et~al.}(2013)\citenamefont {Torrontegui}, \citenamefont {Ib{\'a}{\~n}ez}, \citenamefont {{Mart{\'i}nez-Garaot}}, \citenamefont {Modugno}, \citenamefont {{del Campo}}, \citenamefont {{Gu{\'e}ry-Odelin}}, \citenamefont {Ruschhaupt}, \citenamefont {Chen},\ and\ \citenamefont {Muga}}]{Torrontegui2013}%
  \BibitemOpen
  \bibfield  {author} {\bibinfo {author} {\bibfnamefont {E.}~\bibnamefont {Torrontegui}}, \bibinfo {author} {\bibfnamefont {S.}~\bibnamefont {Ib{\'a}{\~n}ez}}, \bibinfo {author} {\bibfnamefont {S.}~\bibnamefont {{Mart{\'i}nez-Garaot}}}, \bibinfo {author} {\bibfnamefont {M.}~\bibnamefont {Modugno}}, \bibinfo {author} {\bibfnamefont {A.}~\bibnamefont {{del Campo}}}, \bibinfo {author} {\bibfnamefont {D.}~\bibnamefont {{Gu{\'e}ry-Odelin}}}, \bibinfo {author} {\bibfnamefont {A.}~\bibnamefont {Ruschhaupt}}, \bibinfo {author} {\bibfnamefont {X.}~\bibnamefont {Chen}},\ and\ \bibinfo {author} {\bibfnamefont {J.~G.}\ \bibnamefont {Muga}},\ }in\ \href {https://doi.org/10.1016/B978-0-12-408090-4.00002-5} {\emph {\bibinfo {booktitle} {Advances {{In Atomic}}, {{Molecular}}, and {{Optical Physics}}}}},\ \bibinfo {series} {Advances in {{Atomic}}, {{Molecular}}, and {{Optical Physics}}}, Vol.~\bibinfo {volume} {62},\ \bibinfo {editor} {edited by\ \bibinfo {editor} {\bibfnamefont {E.}~\bibnamefont {Arimondo}}, \bibinfo
  {editor} {\bibfnamefont {P.~R.}\ \bibnamefont {Berman}},\ and\ \bibinfo {editor} {\bibfnamefont {C.~C.}\ \bibnamefont {Lin}}}\ (\bibinfo  {publisher} {Academic Press},\ \bibinfo {year} {2013})\ pp.\ \bibinfo {pages} {117--169}\BibitemShut {NoStop}%
\bibitem [{\citenamefont {Pethick}\ and\ \citenamefont {Smith}(2002)}]{Pethick2002}%
  \BibitemOpen
  \bibfield  {author} {\bibinfo {author} {\bibfnamefont {C.}~\bibnamefont {Pethick}}\ and\ \bibinfo {author} {\bibfnamefont {H.}~\bibnamefont {Smith}},\ }\href@noop {} {\emph {\bibinfo {title} {Bose-{{Einstein}} Condensation in Dilute Gases}}}\ (\bibinfo  {publisher} {Cambridge University Press},\ \bibinfo {address} {Cambridge ; New York},\ \bibinfo {year} {2002})\BibitemShut {NoStop}%
\bibitem [{\citenamefont {Schneider}\ \emph {et~al.}(2024)\citenamefont {Schneider} \emph {et~al.}}]{Schneider2024}%
  \BibitemOpen
  \bibfield  {author} {\bibinfo {author} {\bibfnamefont {P.~I.}\ \bibnamefont {Schneider}} \emph {et~al.},\ }\href@noop {} {\bibinfo {title} {In preparation}} (\bibinfo {year} {2024})\BibitemShut {NoStop}%
\bibitem [{\citenamefont {Gardner}\ \emph {et~al.}(2014)\citenamefont {Gardner}, \citenamefont {Kusner},\ and\ \citenamefont {Jake}}]{Gardner2014}%
  \BibitemOpen
  \bibfield  {author} {\bibinfo {author} {\bibfnamefont {J.~R.}\ \bibnamefont {Gardner}}, \bibinfo {author} {\bibfnamefont {M.~J.}\ \bibnamefont {Kusner}},\ and\ \bibinfo {author} {\bibfnamefont {G.}~\bibnamefont {Jake}},\ }\href@noop {} {\bibfield  {journal} {\bibinfo  {journal} {ICML}\ }\textbf {\bibinfo {volume} {2014}},\ \bibinfo {pages} {937} (\bibinfo {year} {2014})}\BibitemShut {NoStop}%
\bibitem [{\citenamefont {Makhlin}\ \emph {et~al.}(2001)\citenamefont {Makhlin}, \citenamefont {Sch{\"o}n},\ and\ \citenamefont {Shnirman}}]{Makhlin2001}%
  \BibitemOpen
  \bibfield  {author} {\bibinfo {author} {\bibfnamefont {Y.}~\bibnamefont {Makhlin}}, \bibinfo {author} {\bibfnamefont {G.}~\bibnamefont {Sch{\"o}n}},\ and\ \bibinfo {author} {\bibfnamefont {A.}~\bibnamefont {Shnirman}},\ }\href {https://doi.org/10.1103/RevModPhys.73.357} {\bibfield  {journal} {\bibinfo  {journal} {Reviews of Modern Physics}\ }\textbf {\bibinfo {volume} {73}},\ \bibinfo {pages} {357} (\bibinfo {year} {2001})}\BibitemShut {NoStop}%
\bibitem [{\citenamefont {Carr}\ \emph {et~al.}(2009)\citenamefont {Carr}, \citenamefont {DeMille}, \citenamefont {Krems},\ and\ \citenamefont {Ye}}]{Carr2009}%
  \BibitemOpen
  \bibfield  {author} {\bibinfo {author} {\bibfnamefont {L.~D.}\ \bibnamefont {Carr}}, \bibinfo {author} {\bibfnamefont {D.}~\bibnamefont {DeMille}}, \bibinfo {author} {\bibfnamefont {R.~V.}\ \bibnamefont {Krems}},\ and\ \bibinfo {author} {\bibfnamefont {J.}~\bibnamefont {Ye}},\ }\href {https://doi.org/10.1088/1367-2630/11/5/055049} {\bibfield  {journal} {\bibinfo  {journal} {New Journal of Physics}\ }\textbf {\bibinfo {volume} {11}},\ \bibinfo {pages} {055049} (\bibinfo {year} {2009})}\BibitemShut {NoStop}%
\end{thebibliography}%
\end{document}